\documentstyle[aps,pre,twocolumn,epsfig]{revtex}

\begin{document}
    \draft

\twocolumn[\hsize\textwidth\columnwidth\hsize\csname @twocolumnfalse\endcsname%

     \title{Diffusion entropy and waiting time statistics of hard x-ray solar flares}
    \author{Paolo Grigolini$^{1,2,3}$, Deborah Leddon$^{1}$, Nicola 
    Scafetta$^{1}$} \address{$^{1}$Center for Nonlinear Science, University 
    of North Texas, P.O. Box 305370, Denton, Texas 76203 }
    \address{$^{2}$Dipartimento di Fisica dell'Universit\`{a} di 
    Pisa, Piazza Torricelli 2, 56127 Pisa, Italy}
    \address{$^{3}$Istituto di Biofisica del Consiglio Nazionale 
    delle Ricerche, Via San Lorenzo 26, 56127 Pisa, Italy }
        \date{\today}
    \maketitle

\begin{abstract}
We analyze the waiting time distribution of time distances $\tau$ between
two nearest-neighbor flares. This analysis is based on the joint use  
of
two distinct techniques.  The first is the direct evaluation of the
distribution function $\psi(\tau)$, or of the probability, $\Psi(tau)$,  
that
no time distance smaller than a given $\tau$ is found. We adopt the
paradigm of the inverse power law behavior, and we focus on the
determination of the inverse power index $\mu$, without ruling out
different asymptotic properties that might be revealed, at larger
scales, with the help of richer statistics. The second technique, 
called
Diffusion Entropy (DE) method, rests on the evaluation of the entropy 
of
the diffusion process generated by the time series. The details of the
diffusion process depend on three different walking rules, which
determine the form and the time duration of the transition to the
scaling regime, as well as the scaling parameter $\delta$. With the first
two rules the information contained in the time series is transmitted,
to a great extent, to the transition, as well as to the scaling regime.
The same information is essentially conveyed, by using the third rules,
into the scaling regime, which, in fact, emerges very quickly after a
fast transition process. We show that the significant information 
hidden
within the time series concerns memory induced by the solar cycle, as
well as the power index $\mu$. The scaling parameter $\delta$ becomes a
simple function of $\mu$, when memory is annihilated. Thus, the three
walking rules yield a unique and precise value of $\mu$ if the memory is
wisely taken under control, or cancelled by shuffling the data. All 
this
makes compelling the conclusion that $\mu = 2.138 \pm 0.01$.
\end{abstract}


]

    \section{Introduction}

The study of solar flares is becoming popular among the researchers working at the frontier of statistical mechanics, due to the widely shared conviction that they are a signature of a significant departure from the  condition of ordinary Brownian motion \cite{parker,Lin,vulpiani,vulpiani2}. As pointed out by Wheatland, \cite{Wheatland}, the
   distribution of times between flares, gives information on how to model flare statistics. In this paper we shall be referring to these times,
   denoted by us with the symbol $\tau$, as the time distance between two consecutive events and the corresponding distribution
   density  will be denoted by $\psi(\tau)$. Although the agreement on the fact that flare statistics depart from ordinary statistical
   mechanics is general, there seems to be the still unsettled issue of what  is the proper model which will account for this form of anomalous
   statistics. Does this form of statistics reflect self-organized criticality or turbulence \cite{vulpiani}?
   We think that the settlement of this delicate issue is made difficult by the fact that, although many authors claim that $\psi(\tau)$ is an
   inverse power law with power index $\mu$, the actual value of $\mu$ still seems to be uncertain. In fact, the authors of Ref. \cite{parker}
   propose $\mu = 1.7$ and those of Ref. \cite{Lin} claim that $\mu = 2$ is the proper  power law index. Boffetta et al. \cite{vulpiani} propose $\mu = 2.4$.
   Finally, Wheatland explains the origin of the power law behavior with a model yielding $\mu = 3.0$, \cite{Wheatland}.

   As it will be made clear by the theoretical analysis of this paper, it is possible to prove, without taking position on the origin of the
   inverse power law behavior, that $\mu = 3$ and $\mu = 2$ 
are critical values. In fact, we note that moving from $\mu > 3$ to $\mu < 3$ is equivalent to a phase transition from the Gaussian to the L\'{e}vy basin of attraction \cite{Annunziato}, and moving from $\mu > 2$ to $\mu < 2$ implies a transition from the  condition of L\'{e}vy statistics to a form of out of equilibrium regime \cite{massi}.
Thus, an uncertainty larger than the distance of the border $\mu = 2$ from the border
   $\mu = 3$ is judged by us to be an unsatisfactory condition that might delay the settlement of the issues concerning the complex dynamics
   underlying the waiting time statistics. The main purpose of this paper is to illustrate a statistical method of analysis that yields a reliable
   value for the power index $\mu$. We hope that this result might be useful for the researchers in this interesting field of investigation and
   at the same time might be beneficial in general for all those who are interested in the statistical analysis of time series.

   The outline of the paper is as follows. In Section II we review the method of Diffusion Entropy (DE) that will be a crucial step of the
   statistical analysis done in this paper. Although the method has been applied somewhere else \cite{nicola1,giacomo,nicvit}, we will present a  short review in order  to make this
   paper as self-contained as possible. In Section III, we illustrate a dynamical model that in general results in time sequences that are
   statistically equivalent to those observed in real data. This model is not limited to the case of inverse power laws, but here we make the
   assumption that the shifted inverse power law is an ideal condition convenient to analyze solar flares, and we study the explicit form emerging
   from this condition. In Section IV, we illustrate two walking prescriptions that will be used to convert the real data into  random
   trajectories. The benefit of adopting several walking prescriptions was discussed in Ref. \cite{giacomo}. Here we introduce two new rules and we apply both of them as
well as one of those introduced in Ref. \cite{giacomo}. 
In Section V we prove that the numerical evaluation of the probability of getting a waiting time larger than a given $\tau$ yields a value for $\mu$ more accurate than that afforded by the waiting-time distribution $\psi(\tau)$. In Section VI we show how to process the data to make an efficient use of the DE method.  In Section VII we use the DE method  to further reduce the error of Section V. We devote Section VIII to concluding remarks.

    \section{Diffusion Entropy}
    The main idea of this approach to scaling is remarkably simple. 
    Let us consider a sequence of $M$ numbers, $ \xi_{i}(t)$ , with  $i = 1, 
    \ldots , M$.     The purpose of the DE algorithm is to establish the 
    possible     existence of a scaling, either normal or anomalous, in the 
    most     efficient way as possible without altering the data with any 
    form     of detrending. Let us select first of all an integer number
        $l$, fitting the condition $1 \leq l \leq M$.       This integer 
    number will be referred to by us as ``time''. For any given     time 
    $l$ we can find $M - l +1$ sub-sequences defined by     \begin{equation}
                \xi_{i}^{(s)} \equiv \xi_{i + s}, \quad   \quad s = 0,  
    \ldots ,  M-l.      \label{multiplicationofsequence}
                \end{equation}          For any of these sub-sequences we 
    build up a diffusion        trajectory, labeled with the index 
                $s$, defined by the position            
                \begin{equation}    x^{(s)}(l) = \sum_{i = 1}^{l} 
    \xi_{i}^{(s)}   = \sum_{i = 1}^{l} \xi_{i+s}.   
                \label{positions}       \end{equation}

                Let us imagine this position as referring to a Brownian 
    particle that      at regular intervals of time has been jumping 
    forward or         backward according to the prescription of the 
    corresponding      sub-sequence of Eq.(\ref{multiplicationofsequence}). 
    This means         that the particle before reaching the position that 
    it holds at        time $l$ has been making $l$ jumps. The jump made at 
    the        $i$-th step has the intensity $|\xi_{i}^{(s)}|$ and is 
    forward or         backward according to whether the number 
    $\xi_{i}^{(s)}$ is         positive or negative. 
                       
               We are now ready to evaluate the entropy of this diffusion 
    process.           To do that we have to partition the $x$-axis into 
    cells of size      $\epsilon(l)$. When this partition is made we have 
    to label the       cells. We count how many particles are found in the 
    same cell at a     given time $l$. We denote this number by $N_{i}(l)$. 
    Then       we use this number to determine the probability that a 
    particle           can be found in the $i$-th cell at time $l$, 
    $p_{i}(l)$, by means       of
               \begin{equation}         p_{i}(l) \equiv  
    \frac{N_{i}(l)}{(M-l+1)} .          \label{probability}
                \end{equation}          At this stage the entropy of the 
    diffusion process at  time  $l$           is determined and reads
                \begin{equation}            S_{d}(l) = - \sum_{i} p_{i}(l) 
    ln [p_{i}(l)].          \label{entropy}
                    \end{equation}         
 The easiest way to proceed with 
    the choice of the cell size, $\epsilon(l)$, is to assume it to be independent 
    of $l$ and determined by a suitable fraction of the square root of the 
    variance of the fluctuation $\xi(i)$. In the case in which the numbers $\xi_i$ are +1, 0 and -1, $\epsilon=1$ is the natural choice. 

             Before proceeding with the illustration of how the DE method 
    works, it is worth making a comment on how to define the 
    trajectories. The method we are adopting is based on the idea of a 
    moving window of size $l$ that 
    makes the $s-th$ trajectory closely correlated to the next, the 
    $(s+1)-th$ trajectory. The two trajectories have $l-1$ values in 
    common. 
It is worth making a comparison with
the technique of Detrended
Fluctuation Analysis (DFA) \cite{simon}. The
DFA is a popular method of scaling
analysis, aiming at detecting the
long-range correlations in seemingly
non-stationary time series that in the
last few years has been used in more
than 100 publications \cite{ivanov}. The DFA
is based on non-overlapping
windows, and, consequently,
trajectories with different labels are
totally independent from one another.
The motivation for using overlapping
windows, with the DE method, is
given by our wish to establish a
connection with the  
Kolmogorov-Sinai (KS) entropy
\cite{beck,dorfman}.
 In Section III we shall make further comments on this connection. The KS entropy of a symbolic sequence is evaluated by 
    moving a window of size $l$ along the sequence. Any window position 
    corresponds to a given combination of symbols, and from the frequency 
    of each combination it is possible to derive the Shannon entropy 
    $S(l)$. The KS entropy is given by the asymptotic 
    limit $lim_{l \rightarrow \infty} S(l)/l$. We believe that the same 
    sequence, analyzed with the DE method, at the large values of $l$ where 
    a finite KS entropy shows up, must yield a well defined scaling 
    $\delta$. To realize this correspondence we carry out the determination 
    of the DE by using the same criterion of overlapping windows as that 
    behind the KS entropy.

             Details on how to deal with the transition from the 
             short-time regime, sensitive to the discrete nature of the 
             process under study, to the long-time limit where both space 
             and time can be perceived as continuous, are given in 
             Ref.\cite{nicvit}. Here we make the simplifying assumption
             of considering large enough times as to make the continuous assumption 
             valid.
    In this case, the trajectories, built up with the above illustrated 
    procedure, correspond to the following equation of motion:
            \begin{equation}
        \frac{dx}{dt} = \xi(t) ,     \label{equationofmotion}
        \end{equation}     where $\xi(t)$ denotes the value that the time 
    series under study     gets at the $\l-th$  site of the sequence under 
    study. 
This means that the time $t = l$ (with $l\gg1$) is thought of as a
continuous and that the function $\xi(l)$ is              a function of
this continuous time.
 In this case the Shannon entropy reads
    \begin{equation}     S(t) = - \int_{-\infty}^{\infty} dx \, p(x,t) ln 
    [p(x,t)].     
\label{continuousshannonentropy}
        \end{equation}     
We also assume that
        \begin{equation}    p(x,t) = \frac{1}{t^{\delta(t)}} \, F\left( 
    \frac{x}{t^{\delta(t)}}\right)         
\label{generalizedscaling}
            \end{equation}  
and  that
$F(y)$ maintains its form, namely that
the statistics of the process is
independent of time. Let us plug
Eq.(\ref{generalizedscaling}) into Eq. (\ref{continuousshannonentropy}). Using a simple
algebra, we get:
    \begin{equation}            S(\tau) = A + \delta(\tau) \tau ,
                \label{keyrelation}             \end{equation}
                where         
  \begin{equation}
                    A \equiv -\int_{-\infty}^{\infty} dy \, F(y) \, ln 
    [F(y)]          \label{ainthecontinuouscase}
                    \end{equation}    
      and
                    \begin{equation}             \tau \equiv ln (t) .
                 \label{logarithmictime}         \end{equation}

  The assumptions made to get the
result of Eq. (\ref{keyrelation}) are not correct during
the transition process, and
consequently the DE method can be
used as a reliable way to detect
scaling only in the long-time limit.
The DE can be used however to shed
light into the regime of transition that
is deeply connected with the
foundation itself of statistical
mechanics. According to Khinchin
 \cite{Khinchin} the central limit theorem is
fundamental for the realization of
canonical equilibrium.
 As well known, a process resulting from the
sum of N independent variables yields a Gaussian distribution, provided that N is large and the single variables have a probability distribution with a finite second moment. A physical process  making N increase from values of the order of unity to  values  so large as to fit the prediction of the central limit theorem can be perceived as a transition from the microscopic  to the macroscopic regime, where thermodynamics applies. If the microscopic variables do not have a finite second moment, the ordinary central limit theorem must be replaced by the generalized central limit theorem \cite{Gnedenko} and in the limiting case of $N \rightarrow \infty$ we find L\'{e}vy rather than Gauss statistics. We can generalize the point of view of Khinchin and consider also in this case the process of transition of N from small to large values as a form of transition from the microscopic to the thermodynamic regime.

          Due to the nature of the DE method, the role of N is here played by the ``time" t. The microscopic regime refers to the fluctuation of $\xi_{i}$ and the macroscopic regime corresponds to the fluctuations of the diffusion coordinate $x(t)$. The time evolution of $\delta(t)$ towards the final value, independent of time, reflects the transition from dynamics to thermodynamics. 

 We shall adopt three
different walking rules (see Section
IV). The first two rules are
characterized by an extended regime
of transition from dynamics to
thermodynamics. Notice that the real
data available are finite, thereby
producing saturation effects in the
long-time regime. Consequently, the
region where the ideal scaling shows
up, is an intermediate time region
following the extended initial
transition and preceding the long-time
saturation regime. This has the effect
of reducing the size of the time region
that can be fruitfully used for scaling
detection. As we shall see, this is the
reason why the DE method must be
supplemented by the use of artificial
sequences. The third rule, on the
contrary, yields a fast transition to the
thermodynamic regime, thereby
allowing us to determine the scaling
by a direct use of Eq. (\ref{keyrelation}), with
$\delta(t)$ assuming the time
independent value of the
thermodynamic limit.

    \section{Dynamic model}
    The solar flares analyzed in this paper are perceived as a sequence of 
    events occurring at unpredictable times, $t_{i}$, with $i=1,\ldots, M$, where
    $M$ is the label of the last event considered. We do not take into 
    account the intensity of these events, which will be studied somewhere 
    else. Thus, the most important property for us to study, is the time 
    distribution density, $\psi(\tau)$, with $\tau$ denoting the time 
    distance between two nearest neighbor events, $\tau_{i} \equiv 
    t_{i+1} - t_{i}$. Let us make the assumption that the experimental 
    analysis of the time series yields the form
        \begin{equation}
        \psi(\tau) = (\mu-1)\frac{T^{\mu-1}}{(T+\tau)^{\mu}}.
        \label{inverse}
        \end{equation}

    We make the key assumption  that the numbers $\tau_{i}$ are 
    uncorrelated. As we shall see, the theory of this paper affords also a 
    criterion to assess if this crucial assumption is correct or not.
    Under this key assumption we can build up a dynamic model that is 
    statistically equivalent to the solar dynamics generating the 
    sequence of the $t_{i}$'s. Let us consider the dynamic process:
    \begin{equation}
        \label{dynamicprocess}
        dy/dt = \lambda y^{z},
        \end{equation}
        with $z > 1$.
        Let us imagine that the trajectory $y(t)$ moves within the 
        interval $[0,1]$. Let us assume also that when the trajectory 
        reaches the right border of this interval it is injected back within 
        this interval by means of a random selection of the initial 
        position $y(0)$. The random selection is done by using a random 
        number generator that assigns the same probability to the numbers 
        of the interval $[0,1]$. 
        The connection between the initial condition and the exit time 
        $\tau$ is 
        given by
        \begin{equation}
            y(0) = [1 + (z-1)\lambda \tau ]^{-\frac{1}{z-1}}.
            \label{initialcondition}
            \end{equation}
            This leads immediately to the distribution of Eq.(\ref{inverse})
            with
            \begin{equation}
                \mu = \frac{z}{z-1}
                \label{powerindex}
                \end{equation}
                and
                \begin{equation}
                    \label{capitalt}
                    T = \frac{\mu -1}{\lambda}.
                    \end{equation}
                    
    We are now equipped to establish a connection with the entropy 
    production per unit of time. Randomness here is involved at the moment 
    of selecting the initial condition, and is characterized by an unknown 
    amount of entropy increase, $H$. If $\mu > 2$, the distribution of 
    Eq.(\ref{inverse}) yields a finite mean waiting time
    \begin{equation}
        <\tau> = \frac{T}{\mu - 2}~.
        \label{ordinary}
        \end{equation}
        It is evident then that the rate of entropy production per unit 
        of time is given 
        by
        \begin{equation}
            h_{E} = H \frac{(\mu -2)}{T}.
            \label{entropyproduction}
            \end{equation}
            Using the dynamic model it is possible to establish a more proper 
            connection with the KS entropy. However, this is of no great 
            relevance within the context of the present paper. 
Therefore, we limit ourselves to considering the entropy production
  of Eq. (\ref{entropyproduction}), where the subscript E stands for "external". In fact, in
  the picture adopted in the present paper the source of entropy
  production is the random selection of the numbers of the interval
  [0,1], an action external to the process under study. It has
  to be pointed out that this external entropy production is subtly
  related to the KS entropy, which, on the contrary, is interpreted as
  being of internal origin \cite{pala}. This is so because the
  dynamical model is a map with a very sharp chaotic region that
  reduces to a set of zero measure, confined to the point $y =1$, in
  the limiting condition where the idealized model of this section
  applies.

            In the case $\mu < 2$ the entropy 
            produced  is proven\cite{massi} to be the following function of time:
            \begin{equation}
                S(t) \propto t^{\mu-1}~.
                \label{nonstationaryentropy}
                \end{equation}              
                It is evident that in the limiting case of very large time values 
                the entropy production per unit of time vanishes, thereby 
                implying that the condition, $\mu = 2$ is a border at which a kind 
                of phase transition occurs. In the region $\mu > 2$ the dynamical 
                system of Eq. (\ref{dynamicprocess}) has an invariant 
                distribution. In the region $\mu < 2$ the system does not have an 
                invariant distribution \cite{massi}. From an intuitive point of view we can 
                imagine that during the observation process the 
                system keeps moving towards an equilibrium distribution, as a kind 
                of Dirac delta function located at $y = 0$ \cite{massi}. The time necessary to reach 
                this invariant distribution is infinite. 

In conclusion, an infinitesimally small change from $\mu > 2$ to $\mu < 2$ would have the effect of annihilating the invariant distribution and of making the process ``non-stationary." The method of analysis of this paper will allow us to assess that $\mu = 2.138 \pm0.01$, namely, that the  solar flares fluctuations are stationary, even if very close to the border with the ``non-stationary" region. This result will be obtained by a direct evaluation of $\mu$, supplemented by the adoption of the DE method. As we shall see, this conclusion is reached after settling a major problem caused by the existence of a genuine form of non-stationary behavior, a fact that is more properly related to dynamic rules changing upon change of time. This will lead us to the final conclusion that the model of Eq. (\ref{dynamicprocess}) is a fairly accurate way of mimicking solar flare dynamics, with $z < 2$ ($\mu > 2$). In Sections VII and VIII we shall make some conjectures on how to
improve this model to take into account the time dependence of the solar
flare rate.

                \section{On three distinct prescriptions to walk}

       The scaling detected by the DE
method is not independent of the
walking rules that we adopt. The
outcomes of DE method are not
unique, due to the dependence of the
scaling parameter $\delta$ on the
walking rules, and this  casts doubts
on this method of analysis. However,
the task of this analysis is an indirect
evaluation of the waiting time
distribution $\psi(\tau)$, or,
equivalently, in the inverse power law
case, of the index $\mu$. If we take for
granted the inverse power law
structure of $\psi(\tau)$, the power
index $\mu$ is unique.
            We adopt the following 
prescriptions for the random walker:

        Rule No. 1.  Make a jump of
fixed intensity, only when you meet
an event, and do it always in the same
direction.
             
Rule No. 2. As with Rule No.
1, make a jump only when you meet
an event, but do it either in the
positive or negative direction
according to a coin tossing
prescription.
            
 Rule No. 3. Walk at fixed
interval of times, with jumps in the
same direction, of intensity
proportional to the time distance
between two nearest-neighbor events.

           Note that here we analyze the                               
sequence $\{\tau_{i}\}$, where each
value $\tau_{i}$ denotes the time
distance between two
nearest-neighbor flares (regarded as
events). Thus, rules No. 1 and No. 2
imply that the random walker makes
instantaneous jumps at the times of
flare occurrence. With rule No. 3 the
random walker, at times $t
=1,2,....,n,....$, makes jumps ahead of
intensity equal to the values $\tau_{i}$
of the sequence under study. Note that
rule No. 1 is one of the two rules used
in Ref. \cite{giacomo}. Here we use for the first
time rules No. 2 and No. 3. Using the
theory of Ref. \cite{giacomo}, which, in turn,
essentially rests on the generalized
central limit theorem \cite{Gnedenko} and on the
work of Feller \cite{feller}, we obtain the
following prescriptions:

              \begin{equation} \label{rule1}\delta = \left\{
                  \begin{array} {ll}
                     \mu-1, & ~~~ ~~1< \mu < 2           \\
                     1/(\mu-1), & ~~~ ~~2 < \mu < 3       \\
                     0.5, & ~~~ ~~ \mu > 3,   \\
                    \end{array} \right.
                    \end{equation}

                   \begin{equation} \label{rule2}\delta = \left\{
                    \begin{array} {ll}
                     0.5 (\mu-1), & ~~~ ~~1< \mu <2          \\
                     0.5, & ~~~ ~~ \mu > 2      \\
                    \end{array} \right.
                    \end{equation}
and
  \begin{equation} \label{rule3}
\delta  = 1/(\mu -1)  , ~~~~~  \mu > 1   ,          
\end{equation}
for rules No. 1, No. 2 and No. 3,
respectively.

               Fig. 1 shows clearly that the
adoption of rule No. 1 alone would
yield two distinct possible values for
$\mu$  when $\delta$ gets  values within
the interval [0.5,1]. However, the
joint adoption of this and the other two rules
settles this ambiguity. We also notice
that both rule No. 1 and rule No. 2 reflect
the phase-transition character of the
condition $\mu = 2$, while rule No. 3,
apparently, does not. However, we
see that rule No. 3 for $\mu < 2$ yields a
value of $\delta > 1$, namely a diffusion
process faster than the ballistic
diffusion. This is a consequence of the
non-stationary nature of the condition
$\mu< 2$.

         It is important to stress that
these rules imply that the numbers
$\tau_{i}$ are not correlated.
Furthermore, these rules rest on the
assumption that the asymptotic limit
of $\psi(\tau)$ is an inverse power law
distribution with no truncation.  We
shall see that the DE method is
sensitive to the correlation among the
numbers $\tau_{i}$, and that the
11-year solar cycle is responsible for
that correlation.  As to the truncation
of the inverse power law at the large
distances, this is another delicate
issue worth of some comments.
Laherr\'{e}re and Sornette, \cite{XX},
suggest that the stretched exponential
family might have a theoretical
motivation stronger than the
power-law distribution. On the other
hand, in the intermediate time region
a stretched exponential is
indistinguishable from a power law.
The two proposed fitting functions
become distinguishable one from the
other in the long-time regime, which
is affected by poor statistics.
However, the work of Refs. \cite{Mantegna} and             
\cite{Floriani} show that a truncation of the
power law of $\psi(\tau)$ at large times
yield an ultra-slow convergence to
normal diffusion, with effects that are
beyond the range of observation of the
DE analysis, due to the data statistical
limitation.

            We shall see that both rule
No. 1 and rule No. 2 yield a very slow
transition to the scaling regime. Due
to the statistical limitation of our data,
the scaling regime turns out to be a
relatively short time region between
transition and saturation regime.
Thus,  we shall be forced to carry out
our analysis with the help of artificial
sequences with the same number of
terms as the real data, by fitting the
DE curves produced by the real data
with the DE curves generated by the
artificial sequences. The adoption of
the third rule, on the contrary, yield a
fast transition to the thermodynamic
regime and, consequently, makes it
possible to make a direct evaluation
of $\delta$. In both cases, however, the
physical consequences of a possible
truncation of the inverse power law
are beyond our range of observation.

    \section{Statistical analysis of the real data: $\psi(\tau)$ and $\Psi(\tau)$}

   In this section we plan to derive the waiting time distribution $\psi(\tau)$ directly from the statistical analysis of the real data, the x-rays emitted by solar flares in the case here under study. At first sight, one might think that a direct determination of $\psi(\tau)$ is more convenient than any indirect approach. Actually, it is not so. As mentioned in Section I, we find that the evaluation of the probability of finding no time distance larger than a given $\tau$, denoted by $\Psi(\tau)$, defined by
\begin{equation}\label{Psi}
\Psi(\tau)\equiv \int\limits_{\tau}^{\infty} \psi(t) dt  ,
\end{equation} 
is more convenient than the direct evaluation of $\psi(\tau)$.  In later sections we shall prove a striking property: the evaluation of $\mu$ through the DE method, an approach less direct than the evaluation of $\Psi(\tau)$, is still more efficient.

The data are a set of 7212 hard x-ray peak flaring event times obtained from the BATSE/CGRO (Burst and Transient Source Experiment aboard the Compton Gamma Ray observatory satellite) solar flare catalog list. The data is a nine-year series of events from 1991 to 2000.  If the time $\Delta t$ between two consecutive solar flares is expressed in seconds, the range goes from 45 to 10,000,000 seconds, as shown in Fig. 2.  Fig. 3 shows the rate of solar flares per month from April 1991 to May 2000. The set of data studied here concerns a time period of 9 years,
and, consequently, a large part of the whole 11-year solar cycle.
Fig. 3 shows that during a large portion of this 11-year cycle the
flare rate undergoes big  changes, thereby significantly departing
from the uniform distribution. Furthermore, it is worth remarking
that, as shown by Fig. 4,
the 11-year solar cycle is not a mere harmonic oscillation with the
period of 11 years, but a complex dynamic process with many
components.

 The direct evaluation of the waiting time distribution, $\psi(\tau)$, needs  the data to be distributed over many bins with
the same size.  When only a few
  data are available,  the bin size cannot be too small, and, in turn, the
  adoption of bins of large size can produce incorrect power law
  indices. 
  In proceeding with the direct evaluation of the
  key parameter $\mu$, first of all, we have to adopt a proper
  criterion to determine the size $\Delta_{i}$ of the i-th bin. We note
  that the waiting time distribution is expected to be an inverse power
  law. If we adopted bins of equal size, those corresponding to large
  times would collect a very limited amount of data, thereby resulting
  in a non reliable  evaluation of the frequencies. To bypass this difficulty we
  adopt bin sizes that are constant in the logarithmic scale. 
This means that  $ln (\tau_i) - ln(\tau_{i-1})$, where $\tau_{i}$ and $\tau_{i-1}$ are the middle times of two consecutive bins, is constant. We define the width of the i-th bin  as $\Delta = \tau_{i} - \tau_{i-1}$, thereby making it become an exponentially increasing function of the sequence position, so as to widely
 compensate for the density decrease. In this
  representation the probability density
  $\psi(\tau_{i})$ is expressed by   
\begin{equation}\label{histogram}
\psi (\tau_i ) = \frac{N_i}{N \Delta_i } ~,
\end{equation}
where $N$ is the total number of data points, $N_{i}$ is number of points located within the
  i-th bin, and $\Delta_{i}$, as earlier said, is the width of the i-th bin.

         The fitting is done by using the prescription of a power law of the type of Eq. (\ref{inverse})
\begin{equation}\label{fit1} 
  \psi(\tau)= \frac{A_1}{(T+\tau)^\mu} ~,   
\end{equation}
with $A_1$, $T$ and $\mu$ being three independent fitting
  parameters.
It is worth noting that the normalization condition reduces the three independent parameters to two, as made clear by Eq. (\ref{inverse}), which is a function of only  T and $\mu$. We find it to be more convenient to adopt three independent fitting parameters, with the understood proviso that the departure of $A_{1}$ from the value $(\mu-1)T^{\mu-1}$ can be interpreted as a way to estimate the inaccuracy of the adopted fitting procedure.

The fitting is done by using an implementation of the nonlinear least-squares (NLLS) Marquardt-Levenberg algorithm \cite{NR}. The NLLS algorithm may not give unique values for the fitting parameters. It needs initial guesses for the free parameters and the final results may change or be affected by huge errors.
This fitting procedure yields: $T = 8787$, $\mu = 2.12 \pm 0.32$ and $A_{1} = 31006$. The evaluated value of $A_1$ is not far from the value 29236 that would be required by the normalization condition. However, there are very large errors of the order of $100\%$, with an error on the  parameter $\mu$ of the order of $15\%$, thereby implying $1.80 < \mu < 2.44$. This means that the result of this fitting procedure would prevent us from assessing the important question raised in Section III on whether the process is stationary or non stationary. The large error of this procedure depends upon the initial values assigned to the three fitting parameters, T, $\mu$ and $A_{1}$, whose choice requires a more efficient criterion. It also depends on the fact that there are oscillations around the fitting curve, as clearly illustrated by Fig. 5.

As earlier mentioned several times, a more accurate fitting is obtained using the function $\Psi(\tau)$. Again we do not pay attention to the normalization constraints and we adopt the following fitting function
       \begin{equation}
        \Psi(\tau) = A_2 \left( \frac{1}{T+\tau}\right)^{\mu-1}.
        \label{inverse2}
        \end{equation}
As shown by Fig. 6, the fitting of the real data is now much more accurate than that of Fig. 5. The fitting parameters used are: $A_{2} = 30657 \pm 16590$, $T = 8422 \pm 500$, $\mu = 2.144 \pm 0.05$. This sets on the key parameter $\mu$ the constraint $2.094 < \mu < 2.194$, which has the very attractive property of establishing the stationary nature of the dynamic model behind the solar flares fluctuations. The results of this search for $\mu$, based on the direct evaluation of $\psi(\tau)$ and on the use of $\Psi(\tau)$, are summarized in Table I. We note that the uncertainty interval associated with the use of $\Psi(\tau)$ is contained within  the wider uncertainty interval produced by the use of $\psi(\tau)$. This means that we are coming closer to the real value of $\mu$. The width of the uncertainty interval will be further reduced by using the DE method.

\begin{center}Table I \\
\begin{tabular}{|c|c|}
\hline $\psi(\tau)$ & $\Psi(\tau)$  \\
\hline ~$1.80<\mu<2.44$ & ~$2.094<\mu<2.194$ \\
\hline
\end{tabular} 
\end{center}

\section{Diffusion Entropy of solar flares.}

        This section is devoted to the analysis of the solar flares data by means of the DE method. The final result will be given by $\mu = 2.138 \pm  0.01$, namely a value for $\mu$ even more accurate than that obtained in Section V by using $\Psi(\tau)$. We shall prove also that the DE method allows us to establish some aspects of the dynamics behind solar flares that would be overlooked by an analysis based only on the use of the waiting time distribution.

         The first issue that we have to solve is how to process the data so as to apply the three walking rules of Section IV. The data accessible to us are the times $\tau_{i} = t_{i} - t_{i-1}$, with $t_{i}$ and $t_{i-1}$ denoting the time of occurrence of the i-th and the (i-1)-th solar flare, respectively. However, the direct adoption of these numbers would result in technical difficulties that are bypassed by referring ourselves to the new sequence of numbers
\begin{equation}\label{betadat}
\beta_j= Int\left[\frac{\Delta t_j}{\Lambda}\right] +1,
\end{equation}  
where $Int[x]$ denotes the integer part of $x$.  The adoption of $\Lambda = 1$ would be virtually equivalent to referring ourselves to the original sequence of numbers. However, preliminary trials with changing values of $\Lambda$ led us to conclude that there are problems with the adoption of  both excessively small and excessively large values of $\Lambda$. The adoption of excessively small values of $\Lambda$ would make the computer analysis too slow and would require an excessively large amount of computer memory. This is the reason why we cannot use the original sequence of numbers. The adoption of excessively large values of $\Lambda$, on the other hand, would produce statistical saturation, and a consequent sub-diffusion process that would not accurately reflect the dynamics behind the data. We adopted the criterion of using the largest value of $\Lambda$ compatible with negligible saturation effect. Preliminary attempts made it possible for us to assess that this convenient value is given by $\Lambda = 3600$.

After processing the data, we have to realize the three walking rules of Section IV. We note that according to the prescription of Section II, diffusion is generated by the random walker jumping at any time step. The random walker makes jumps of intensity $|\xi_{i}|$, ahead or backward, according to whether
$\xi_{i}>0$ or $\xi_{i} < 0$.  Thus, we create a new sequence ${\xi_{i}}$, of 0's and 1's, with the following prescription. We consider a sequence of infinite empty sites, labeled by the integer index $i$, considered as a discrete time, running from $i = 1$ to $i = \infty$. We divide this sequence into patches of width $\beta_{j}$. The first patch consists of the sites $i =1$, $i=2, $Š.., $i=\beta_{1}$, the second patch consists of the sites $i = \beta_{1} + 1$,  $\beta_{1} + 2$, Š.., $  \beta_{1}+\beta_{2}$, and so on. We assign the value 0 to all the sites of the same patch but the last site.
This means that the random walker walks only at the end of the patch, namely, at the occurrence time of an event. To apply rule No. 1, with the random walker always moving in the same direction, we always assign to the last site of a given patch the value of 1.
To apply rule No. 2 we assign to the last site of any patch either the value 1 or the value -1, according to the coin tossing rule. The coin tossing prescription is realized by using a random number generator. To reduce the risk of artificial periodicity we create 10 different sequences, each corresponding to a different random distribution of 1's and -1's. For any sequence we run the DE method and then we make the average over the 10 resulting DE curves. To apply the rule No. 2, which will be shown in action in
Section  VII C, we have to identify $\xi_{i}$ with $\beta_{i}$.

        The DE results obtained applying rule No. 1 are illustrated in Fig. 7.
This figure shows one of the benefits of the DE method.
According to rule No. 1, we have to use the prescription of Eq.
(\ref{rule1}). The most accurate of the values of $\mu$, discussed in
Section V, is $\mu = 2.144$. This value, being smaller than 3 and
larger than 2, makes us adopt the formula $\delta = 1/(\mu -1)$,
and yields the scaling parameter $\delta = 0.874$, which is the slope
of the straight line of Fig. 7.

 This theoretical  prediction implies that the
times $\tau_{i}$ of the sequence $\{\tau_{i}\}$ are not correlated with each other.
 In the specific case of seasonal periodicity described by harmonic oscillations, the numerical results of Ref. \cite{nicola1} prove that the scaling detected by the DE, as well as by other methods to detect scaling, is higher than the Brownian motion scaling $\delta = 0.5$. This is so even when there is no correlation in addition to seasonal periodicity. We eliminate this effect, by shuffling the data.  The DE method can be applied to both the original sequence of $\beta_{i}$ and to the shuffled sequence. If the DE yields two different curves, this is a proof of the fact that there is memory in the original sequence. This is an important property that cannot be revealed by the analysis of the waiting time distribution, $\psi(\tau)$.  Fig. 7 shows that this is the case. In fact we see that the DE curve corresponding to the shuffled data, after the transition region at short time and before saturation, has a slope distinctly smaller than the curve referring to the non shuffled data. Furthermore, this slope is closer to the slope of the straight line corresponding to the finding of Section V, which yields $\mu = 2.144$, and, consequently, according to Eq. (\ref{rule1}), $\delta = 0.874$.  However, both shuffled and non-shuffled data yield saturation effects at a time scale of the order of $t_{sat} = 1,500$ hours. These saturation effects set limits to the accuracy of the determination of the value of $\mu$ by means of the DE method.

        In Fig. 8   we illustrate the results obtained by using rule No. 2. It is remarkable that in this case the shuffled data yield, with the DE method, an entropy increase faster (rather than slower) than the non-shuffled data. This is a consequence of the fact that in this case the deviation from ordinary diffusion,  produced by time periodicity, would generate sub-diffusion rather than super-diffusion.  We notice that the difference between the shuffled and non-shuffled curves is smaller than that in the case of Fig. 7 (rule No. 1) and that the saturation effects show up at later times. We thus conclude that rule No. 2 is much less sensitive to periodicities and to saturation effects than rule No. 1.

\section{A FURTHER IMPROVEMENT: USE OF ARTIFICIAL SEQUENCES}

We have seen that the DE method reveals the existence of memory effects that are overlooked by the direct evaluation of the waiting time distribution. However, as pointed out in Section II and illustrated by the numerical results of Section VI, the time region where the DE method might be fruitfully used to detect scaling, is reduced to an intermediate time region, after the transition from dynamics to thermodynamics, and before the saturation effects. This has the unwanted effect of setting limitations to the accuracy of the DE method. To bypass this difficulty we generate artificial sequences with the same statistical limitations of the real data, and then we search for the parameter $\mu$ that establishes the most accurate fitting with the DE curves derived from real data.

        To make this procedure as reliable as possible we proceed as follows. We assume that $\psi(\tau)$ has the form
\begin{equation}\label{fittt1} 
  \psi(\tau)= \frac{A}{(T+\tau)^\mu} ~,   
\end{equation}
where $T$  and $\mu$ are our fitting parameters. The constant A is determined by the normalization condition through
\begin{equation}\label{coefA}
  \frac{1}{A} \equiv  \int_{45}^{\infty } \frac{1}{(T+\tau)^\mu} d\tau ~.
\end{equation}
The fitting parameters are made to change around the mean values established by the results of Section V which yield $\mu = 2.144 \pm 0.05$ and $T = 8422 \pm 500$. Note that in the real data no time exists with a value smaller than $\tau = 45$ sec.
 This is the reason why the integration in Eq. (28) is done from $45$ to $\infty$ rather than from $0$ to $\infty$.  The number of data available to us are 7211. Thus we produce 7211 values of $\tau_{i}$, 
according to the prescription
\begin{equation}\label{artgen}
\tau_{i}=\left[\frac{1}{\left(T+45 \right)^{\mu-1}} -  \frac{(\mu-1)~y_i}{A} \right] - T  ~,
\end{equation}
with the number $y_{i}$ randomly selected in the interval $[0,1]$. It is straightforward to prove that the resulting distribution of $\tau_{i}$
is the same as that of Eq. (\ref{fittt1}) and fits the condition of Eq. (\ref{coefA}).
At this stage we are ready to compare the DE curves generated by the artificial data to the DE curves generated by the real data, using both rule No. 1 and rule No. 2. The comparison is made with the DE curves corresponding to shuffled data, since the artificial sequences are generated without correlation among the numbers $\tau_{i}$.

Let us discuss first the results concerning rule No. 1. These results are illustrated in Figs. 9. In Fig. 9a we show the effect of changing $\mu$ in the interval $[2.094, 2.194]$, with $T = 8422$ and in Fig. 9b we show the effect of changing $T$ in the interval $[7922, 8922]$, with $\mu = 2.144$. We see that the DE curves of the artificial sequences fluctuate within an error strip containing the DE curve of the real data.
 The size of this error strip increases upon change of time and we see that the spreading caused by the change of $T$ is much smaller than that caused by the change of $\mu$.
From a qualitative point of view, the 
results concerning rule No. 2, shown in Figs. 10a and 10b, are very similar.

\subsection {A  more accurate measurement of $\mu$.}

 We have seen that the area of the T-error strip is significantly smaller than that of the $\mu$-error strip, at least five times smaller. Therefore, we can improve the accuracy of $\mu$ by assigning to $T$ a fixed value and looking for the value of $\mu$ ensuring the best fitting of the real data. We assign to $T$ the value of $8422$, and we proceed with the search for the best fitting. The results are illustrated in Figs. 11a and 11b. The result concerning rule No. 1 is good, as seen in  Fig. 11a.  As expected, Fig. 11b shows that the result concerning rule No. 2 is even better, and we think that it can be judged to be excellent. This extremely accurate result is due to the DE curve of the artificial sequence coinciding with the DE curve of real data over the wide range of $1000$ hours of diffusion. On the basis of this excellent fitting, we conclude that 
\begin{equation}\label{bestfitmu}
\mu = 2.138 \pm 0.01 ~.
\end{equation}

\subsection{Non shuffled data and an artificial sequence
 with suitable memory. }

        In Section VI, we have noticed that the result of the DE analysis depends on whether the real data are shuffled or not. We think that in the original data there are signs of the 11-year solar cycle and other subcycles. This makes it harder to establish a connection between the scaling
$\delta$ and the power index $\mu$. However, if our conclusion that $\mu = 2.138 \pm 0.01$ is correct, it should be possible to fit the DE curve of the non-shuffled original data with no further change of the fitting parameters T and $\mu$, provided that we sort the  artificial sequence in such a way as to mimic the solar periodicity. Rather than doing that with a model, for instance a suitable modulation of the parameter $\lambda$ of Eq. (12), we proceed in a more direct way, according to the following procedure. Let us call $R_{i}$ and $A_{i}$ the $i-th$ numbers   of the real and artificial sequence used in subsection A, respectively.  The $i-th$  number of the sorted artificial sequence is denoted by $S_{i}$. The subscript $i$ ranges from 1 to N. The number $S_{1}$ is fixed by selecting from the set of $A_{i}$'s the number that is closest to $R_{1}$, this being, let us say,  $A_{j(1)}$. We thus set $S_{1} = A_{j(1)}$. The number $A_{j(1)}$ is eliminated from the artificial sequence. Then, we move to $R_{2}$ and from the set of the remaining N-1 numbers of the artificial sequence we select the closest one to it, this being, let us say, $A_{j(2)}$. We proceed with the same criterion until we exhaust all the numbers of the artificial sequence. It is evident that the adoption of this procedure assigns to the
artificial data a time order reflecting the complex dynamics
illustrated by Figs. 2 and 3.

        At this stage, we evaluate the corresponding DE curve and we compare it to the DE curve generated by the non-shuffled real data. As earlier mentioned, the sorted artificial data are the same as those used to produce the excellent fitting of the DE curves derived from the shuffled original data. Thus,  the  fitting parameters are the same as those used for Figs. 11. We illustrate the new result  in Figs. 12, which show that the fitting accuracy is as good as (and for rule No. 1 even slightly better than) the fitting of Figs. 11. This is a very remarkable result since Figs. 7 and 8 show that shuffling the data produces a significant effect. Thus, Figs. 11 and 12 prove that the memory of the data is totally under our control.

\subsection{Third rule in action.}

According to Lepreti, Carbone and
Veltri \cite{Lepreti} the waiting time
distribution $\psi(\tau)$ is already
L\'{e}vy.  This would imply that the
adoption of the third rule yields an
infinitely fast transition from
dynamics to thermodynamics. This is
so because L\'{e}vy distribution is
stable and the convolution between
two distinct L\'{e}vy distributions is a
L\'{e}vy distribution \cite{Gnedenko}. According
to our analysis, $\psi(\tau)$ is a shifted
inverse power law. It is plausible that
the difference between the shifted
power law distribution of Fig. 5 and
the L\'{e}vy distribution of Ref. \cite{Lepreti}
is small. Consequently, the transition
to thermodynamics is expected to be
very fast. This expectation is
confirmed by the  numerical results
illustrated in Fig. 13. The transition to
the scaling regime is so fast that it is
possible to detect a wide regime of
linear dependence of the entropy on
$log(l)$, which allows us to derive for
$\mu$ the value   $\mu = 2.138$, in total
agreement with the conclusion of the
earlier analysis done by means of
rules No. 1 and No. 2. We see that in
this case the memory of the
non-shuffled data yields a $\delta$
slightly larger than the scaling
parameter of the shuffled data.   The
adoption of rule No. 3
implies a statistical accuracy smaller
than that of the other two rules, due to
fact  there is no limitation to the 
jumps intensities, thereby decreasing
the number of particles located in the
same cell. This has the effect of
making the evaluation of $p_{i}$ and
consequently that of the entropy less
accurate.  However, this disadvantage
is widely compensated by the
emergence of a much more extended
scaling region that yields as a result a
value of $\mu$  fully confirming that of
the other two rules.

    \section{concluding remarks}

        We see that the uncertainty on the value of  $\mu$ for solar flares has been significantly reduced. The current literature, if we give the same credit to all the authors, yields values of $\mu$ ranging from 3 to 1.7. We  provide  the compelling conclusion that $\mu = 2.138 \pm 0.01$. However, this is not the main result of his paper. We think that this paper shows that the DE method is a remarkably accurate technique of analysis that goes much beyond the direct evaluation of the waiting time distribution $\psi(\tau)$. This is so because complex processes are characterized by two different kinds of memory.  The memory of first kind is the main object of the research work done in the field of the Science of Complexity. To make clear the nature of this kind of memory, let us recall \cite{Bedeaux} that a Markov master equation, namely a stochastic process without memory, is characterized by a waiting time distribution $\psi(\tau)$ with an exponential form, thereby implying memory for a marked deviation from the exponential condition. This is why the search for an inverse power law distribution with a finite value of $\mu$ (the exponential distribution means $\mu = \infty$) can be interpreted as a search for memory. This is the memory of the first kind, to which the prescriptions of Ref. \cite{giacomo} are referred to. For real data, in addition to this form of memory, another type of  memory might be present, denoted by us as memory of the second type, under the form of correlation among the values $\tau_{i}$. In this paper we have seen that this second form of memory is given, in  this case, by the 11-year solar periodicity. It is possible that this form of additional memory is present in many other complex processes for different reasons. It is also evident that it is difficult, or perhaps impossible to reveal this form of additional memory by means of the direct evaluation of $\psi(\tau)$. This paper proves that joint use of the direct evaluation
of $\psi(\tau)$ (or of $\Psi(\tau)$) and of the DE method is a very useful supplement to the ordinary technique, and that it can be profitably used to shed light on the dynamics behind the time series generated by complex processes.

This paper yields a convincing conclusion concerning the distinction between two possible forms of non-stationary behavior. As pointed out in Section III, the claim that the waiting time distribution $\psi(\tau)$ has the form of Eq. (\ref{inverse}) is equivalent to assuming that the dynamics of the flaring process is driven by the model of Eq. (\ref{dynamicprocess}) with the assumption that the  trajectories are injected back randomly. This is a stationary model that in the case where
$z> 2$, $(\mu > 2)$, would be incompatible with the existence of an invariant distribution \cite{massi} and consequently with ``thermodynamic equilibrium". The inaccuracy of the analyses done by the earlier work in this field would prevent us from distinguishing this form of non-stationary behavior from
a genuinely form of non-stationary behavior. By genuinely non-stationary behavior, we mean the existence of rules changing with time. This form of genuinely non-stationary behavior might be modelled, for instance, by assuming that the parameter $\lambda$ of Eq. (\ref{dynamicprocess}) is time dependent. If we make the assumption that the time dependence of $\lambda$ has a period of 11 years, and we make our analysis over a period of time that is not much larger than this time period, as we have done, then the process must be perceived as being genuinely non stationary. Our analysis is so accurate as to rule out the former form of non-stationary behavior and to detect significant effects stemming from the latter, or, equivalently, from the existence of the memory of the second type. 

In this paper we do not take side with either the proponents of self-organized criticality \cite{a} or with those of turbulence \cite{vulpiani,vulpiani2}. The dynamical model of Section III is inspired by the models of turbulence, but we mainly use it to generate artificial sequences mimicking the real ones with no claim that it is an exhaustive picture of the dynamics behind solar flares. The fitting of Fig. 6  seems as good as the fitting of Fig. 1 of Ref. \cite{Lepreti}. However, our analysis does not rest only on the waiting time distribution. In a very recent paper Wheatland \cite{b} criticized the work of Ref. \cite{Lepreti} as being based on the assumption that rate of solar flares is constant. This is not so, as  shown by Fig. 3. On the other hand, modelling the time dependence of this rate is not easy, since it does not correspond only to a 11-year periodic motion but to a much more complex
condition, as illustrated in Fig. 4. In fact, this figure shows
that there are many other components in action. This is the reason why we decided to mimic the time dependence of the solar flare rate sorting the artificial sequence in the way described in Section VII B. We found that this yields a fitting with the real data as good as the fitting between the DE curve produced by the artificial sequence, with no sorting induced memory, and the DE curve produced by the shuffled real data. This is, in our opinion, a strong indication  that the value of $\mu = 2.138$ is a genuine property of real data. On the other hand, the dynamical model of Section III can also be adapted to reproducing the modulated Poisson process advocated by Wheatland \cite{b}. This is left as a subject of future investigation. Even in this case, the role of the DE analysis is expected to be crucial, and the result is expected to strongly depend on wether the modulation process involves randomness or only quasi-periodicity.

        In our notation, the power index found by the authors of Ref. \cite{Lepreti} is  $\mu = 2.38$, a value that turns out to be compatible with the uncertainty interval associated to the determination of $\mu$ by means of the direct evaluation of $\psi(\tau)$. Our analysis establishes a connection with L\'{e}vy statistics, in accordance again with the conclusions of Ref. \cite{Lepreti}.
However, we adopt a perspective that is different from that of the
authors
of Ref. \cite{Lepreti}. Our diffusion process reaches the L\'{e}vy regime
after the process of transition from dynamics to thermodynamics
that has been discussed in detail in the earlier sections. This
process is very fast if the rule No. 3 is adopted, but it is not
infinitely fast as in the perspective of the authors of Ref. \cite{Lepreti}
who assume the waiting time distribution $\psi(\tau)$ to obey
already the L\'{e}vy statistics.  It is worth pointing out that the
perspective adopted in this paper makes it possible to take into
account the time dependence of the solar flare rate.
 We do not rule out the possibility that $\psi(\tau)$ is a stretched exponential \cite{XX}. In fact, a stretched exponential would not conflict with the attainment of L\'{e}vy statistics in the long-time limit of the diffusion process. Although a truncation of $\psi(\tau)$ at large values of $\tau$ generates a finite second moment, and consequently Gaussian statistics in the long-time limit, the transition to the conventional thermodynamic regime is ultra slow \cite{Mantegna}. It is known \cite{Floriani} that a much earlier transition to L\'{e}vy statistics occurs and that the L\'{e}vy regime lasts for a very extended period of time. The transition to the Gaussian regime probably takes place at times much larger than the saturation time, and might be made visible only in the ideal case of infinitely large sequences.
\newline

{\bf Acknowledgements:} we thank the BATSE/CGRO team, NASA/Goddard Space Center,  
Greenbelt, Md, for generously providing the data.

\newpage
\onecolumn

\begin{figure}[h]
\epsfig{file=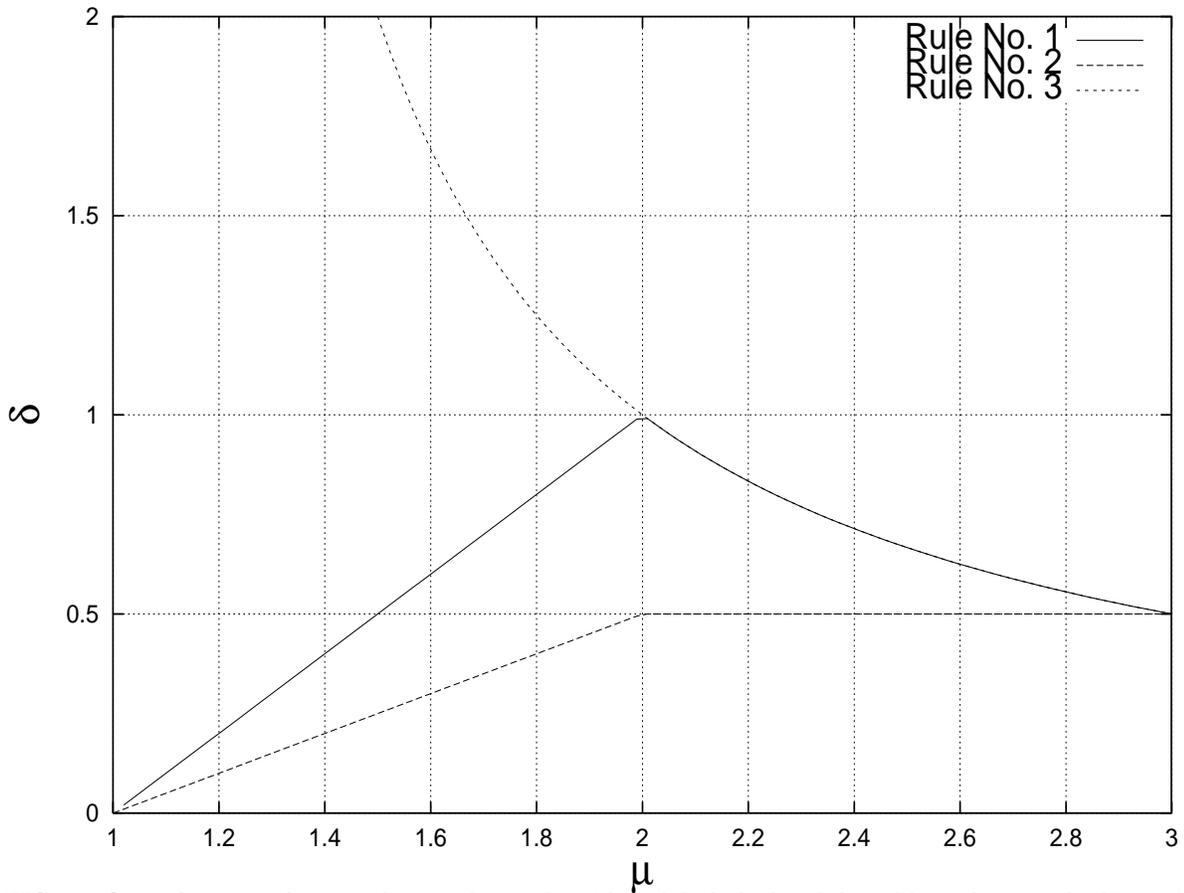, height=16cm,width=12cm,angle=-90}
~\\

\caption{ $\delta$ as a function of $\mu$ according to three rules. The solid,
dashed and dotted lines denote rules No. 1, No. 2 and No. 3,
respectively.    }
\end{figure}

\newpage

\begin{figure}[h]
\epsfig{file=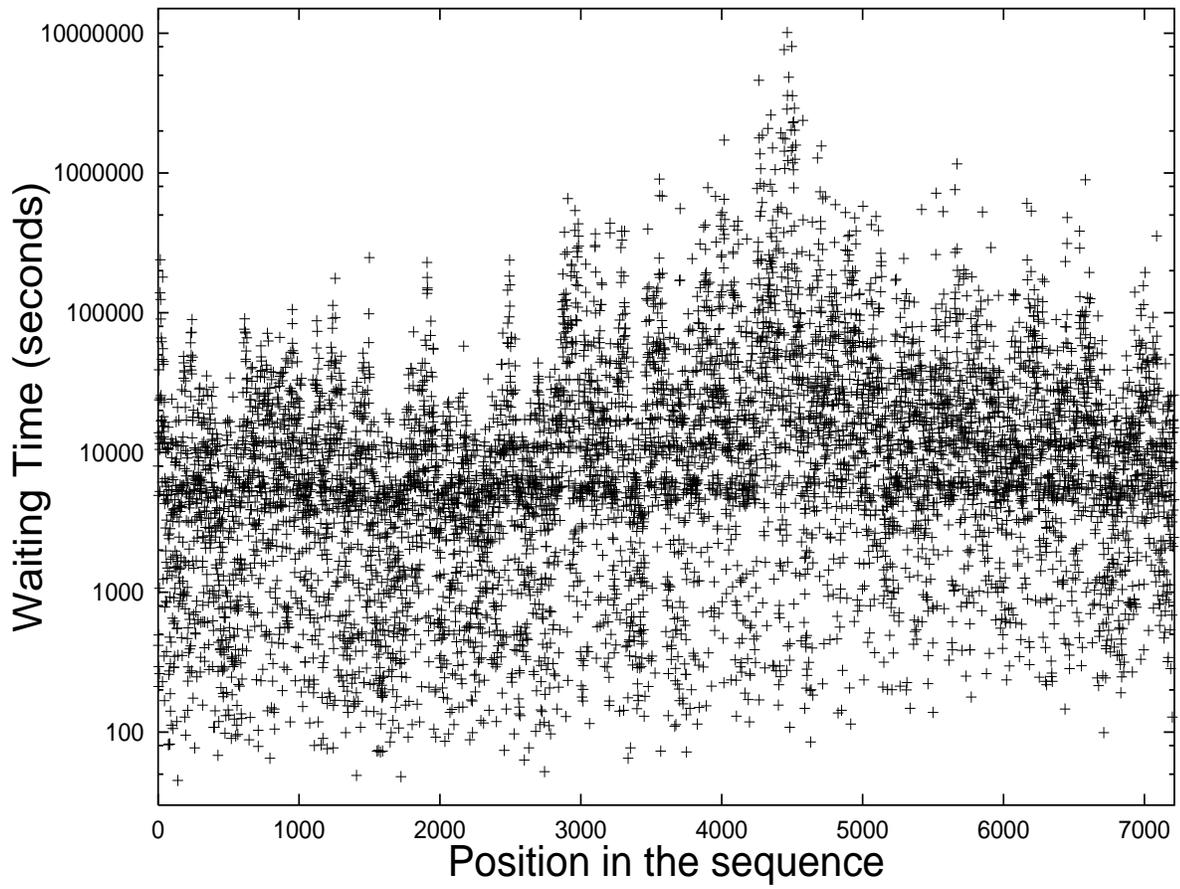, height=16cm,width=12cm,angle=-90}
~\\

\caption{
The original sequence of the solar flares waiting times. Note the logarithmic scale of ordinates.}
\end{figure}

\newpage

\begin{figure}[h]
\epsfig{file=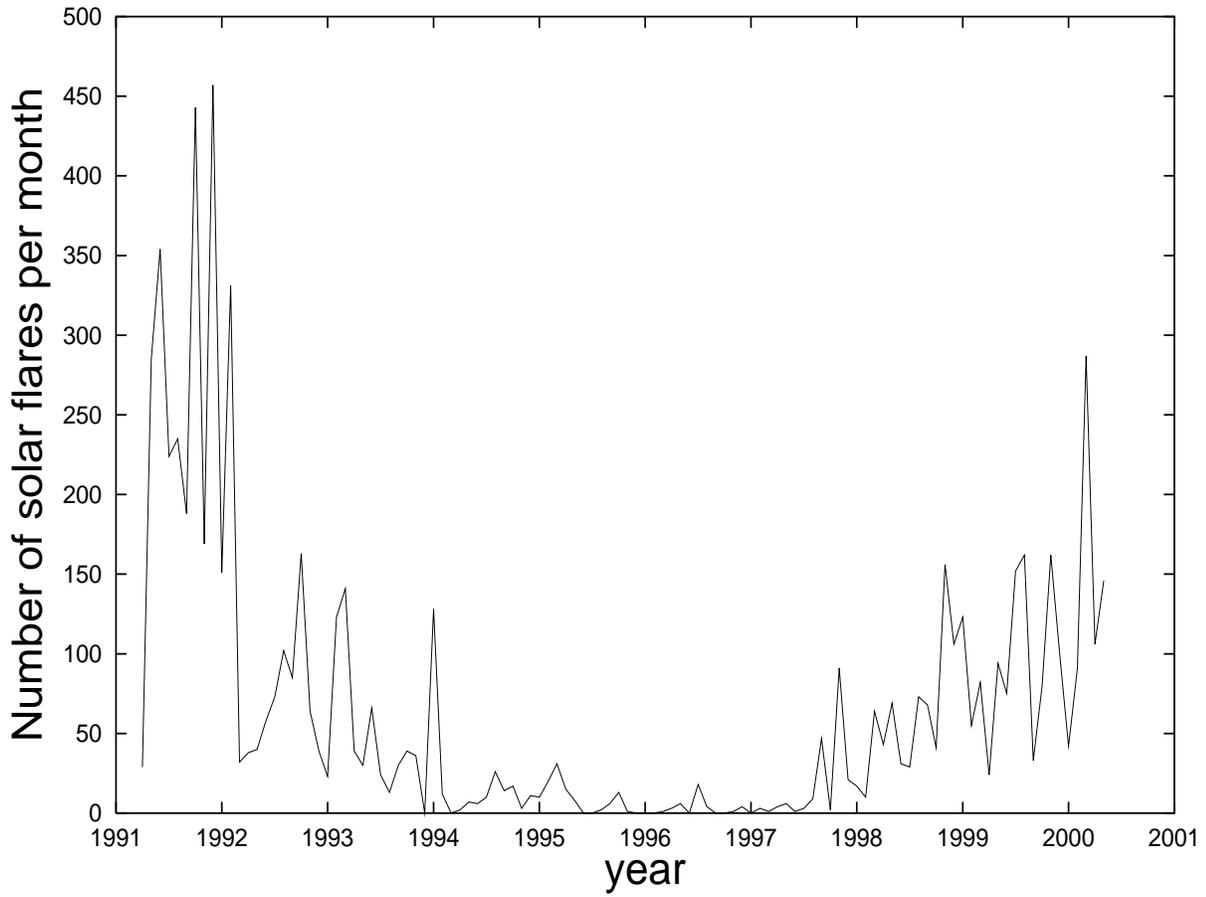, height=16cm,width=12cm,angle=-90}
~\\

\caption{ Number of solar flares per month from April 1991 to May 2000.  }
\end{figure}

\newpage

\begin{figure}[h]
\epsfig{file=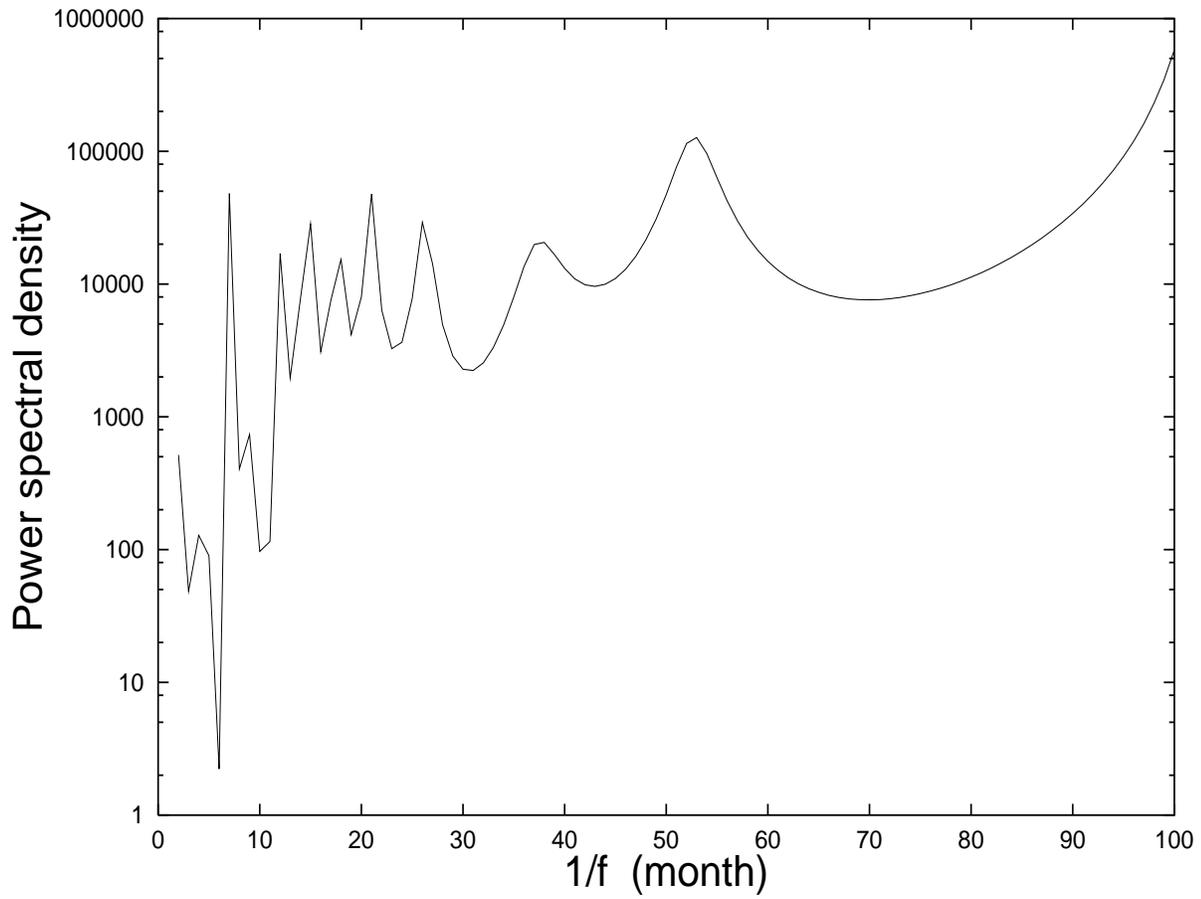, height=16cm,width=12cm,angle=-90}
~\\

\caption{ 
The solid curve was obtained by using the maximum entropy
method [22].}
\end{figure}

\newpage

\begin{figure}[h]
\epsfig{file=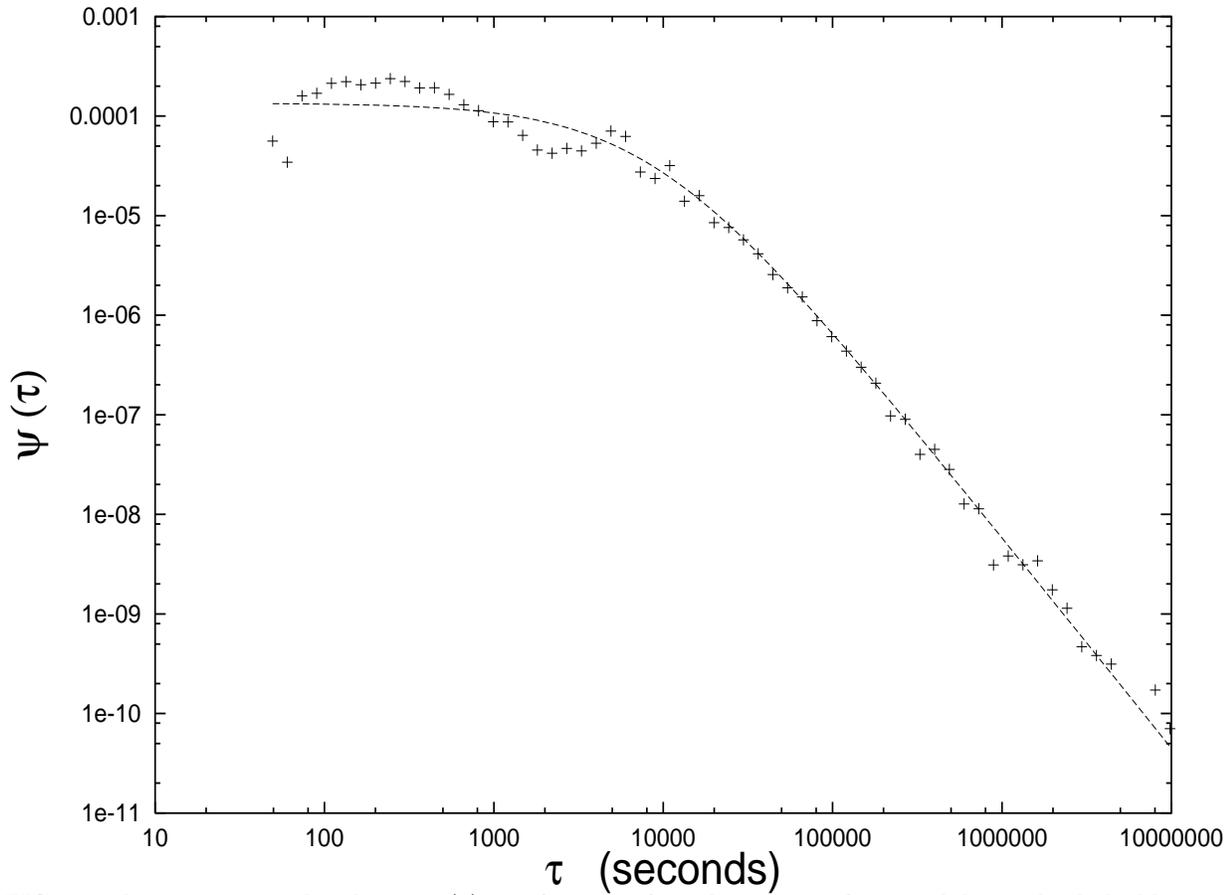, height=16cm,width=12cm,angle=-90}
~\\

\caption{ 
The waiting time distribution $\psi(\tau)$ as a function of $\tau$. The crosses refer to real data. The dashed line is the fitting function of Eq. (\ref{fit1}) with $A_{1} = 31006 $, $T = 8787  $ and $\mu =2.12 $. }
\end{figure}

\newpage

\begin{figure}[h]
\epsfig{file=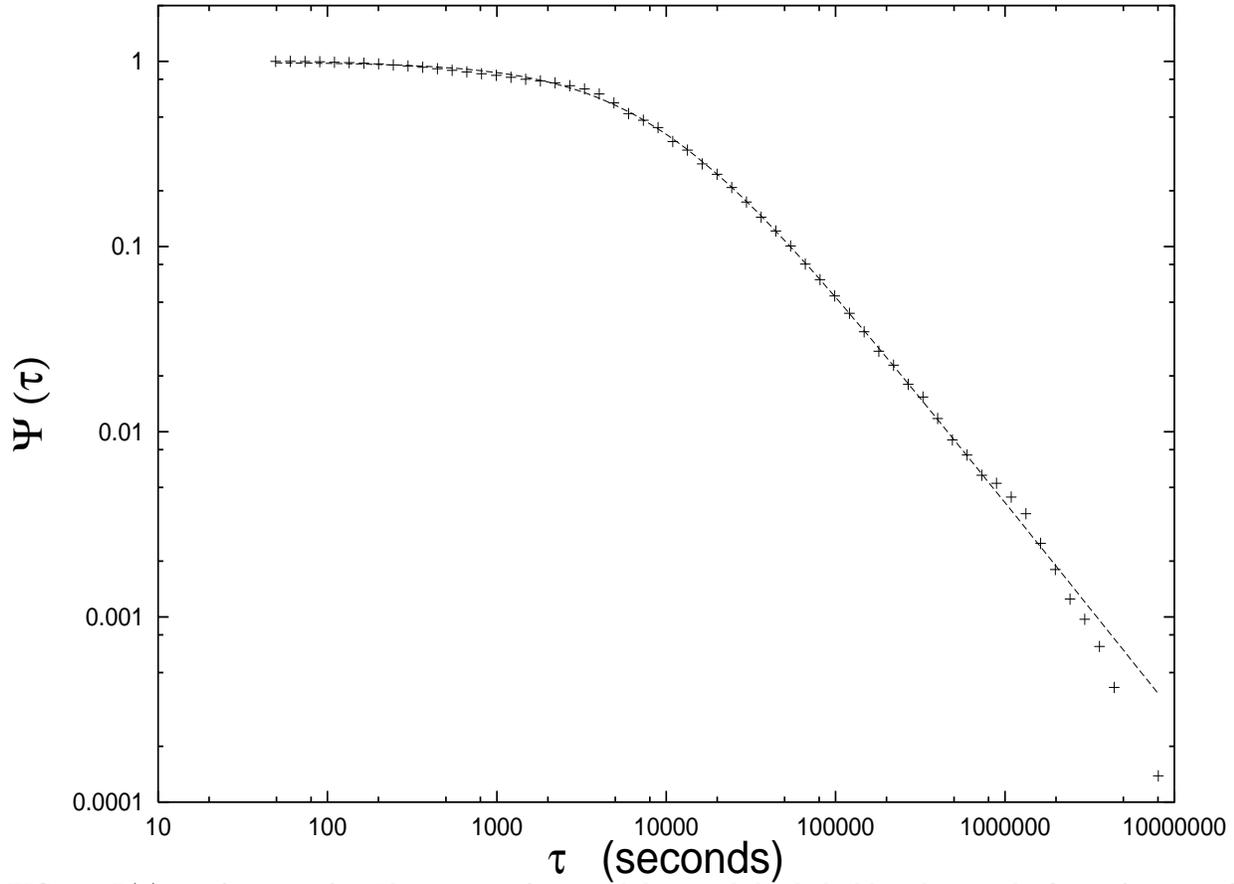, height=16cm,width=12cm,angle=-90}
~\\

\caption{
$\Psi(\tau)$ as a function of $\tau$. The crosses refer to real data, and the dashed line denotes the fitting function of Eq. (\ref{inverse2}) with  $A_{2} = 30567$,  $T = 8422 $  and $\mu = 2.144$.  }
\end{figure}

\newpage

\begin{figure}[h]
\epsfig{file=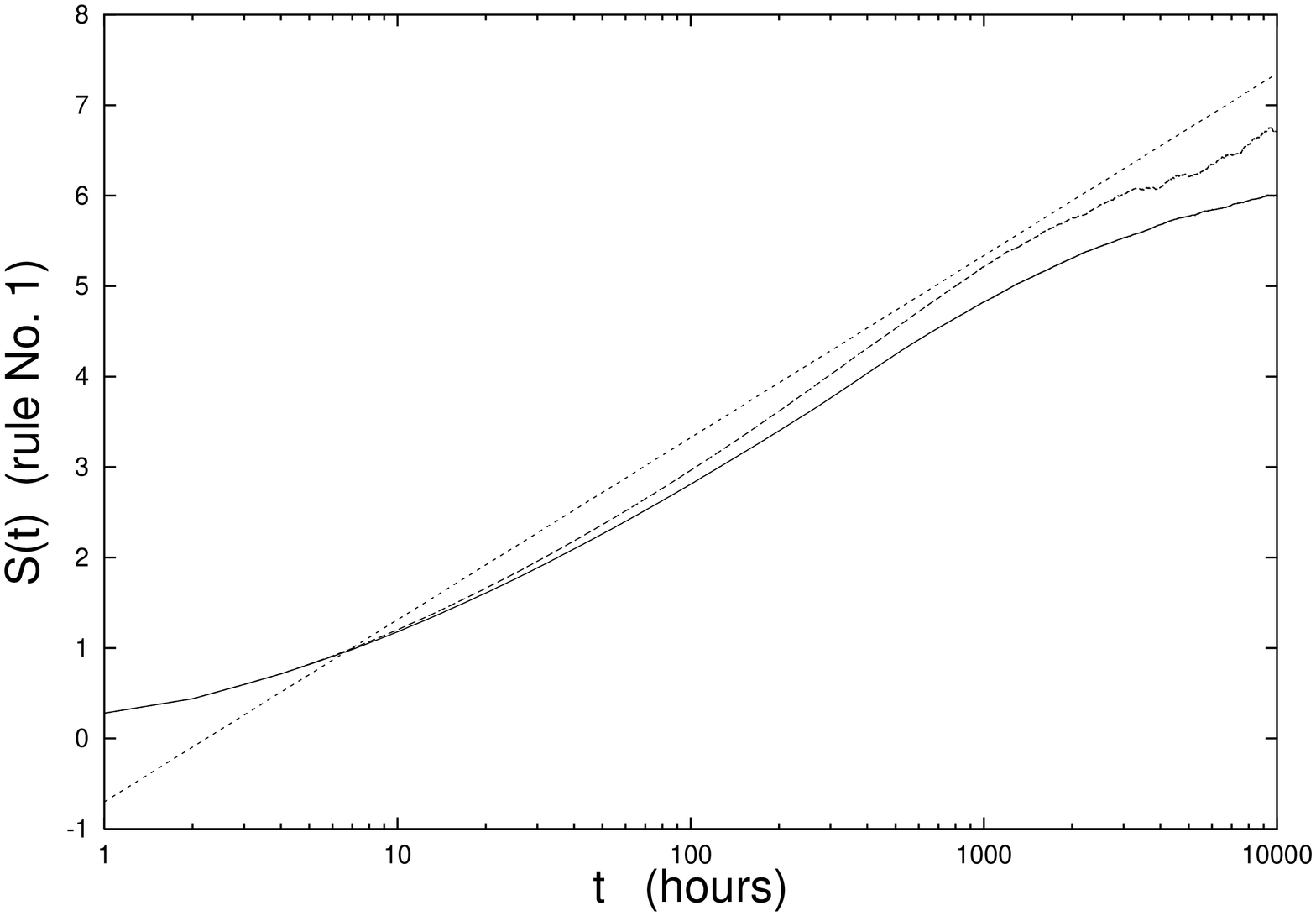, height=16cm,width=12cm,angle=-90}
~\\

\caption{
DE as a function of time according to rule No. 1. The dotted straight line illustrates the slope of entropy increase corresponding to $\mu = 2.144$,  and $\delta=0.874$, which is the best value of $\mu$ afforded by the analysis of Section V. The dashed line is the DE curve generated by the non-shuffled real data. The solid line is the DE curve generated by the shuffled real data.}
\end{figure}

\newpage

\begin{figure}[h]
\epsfig{file=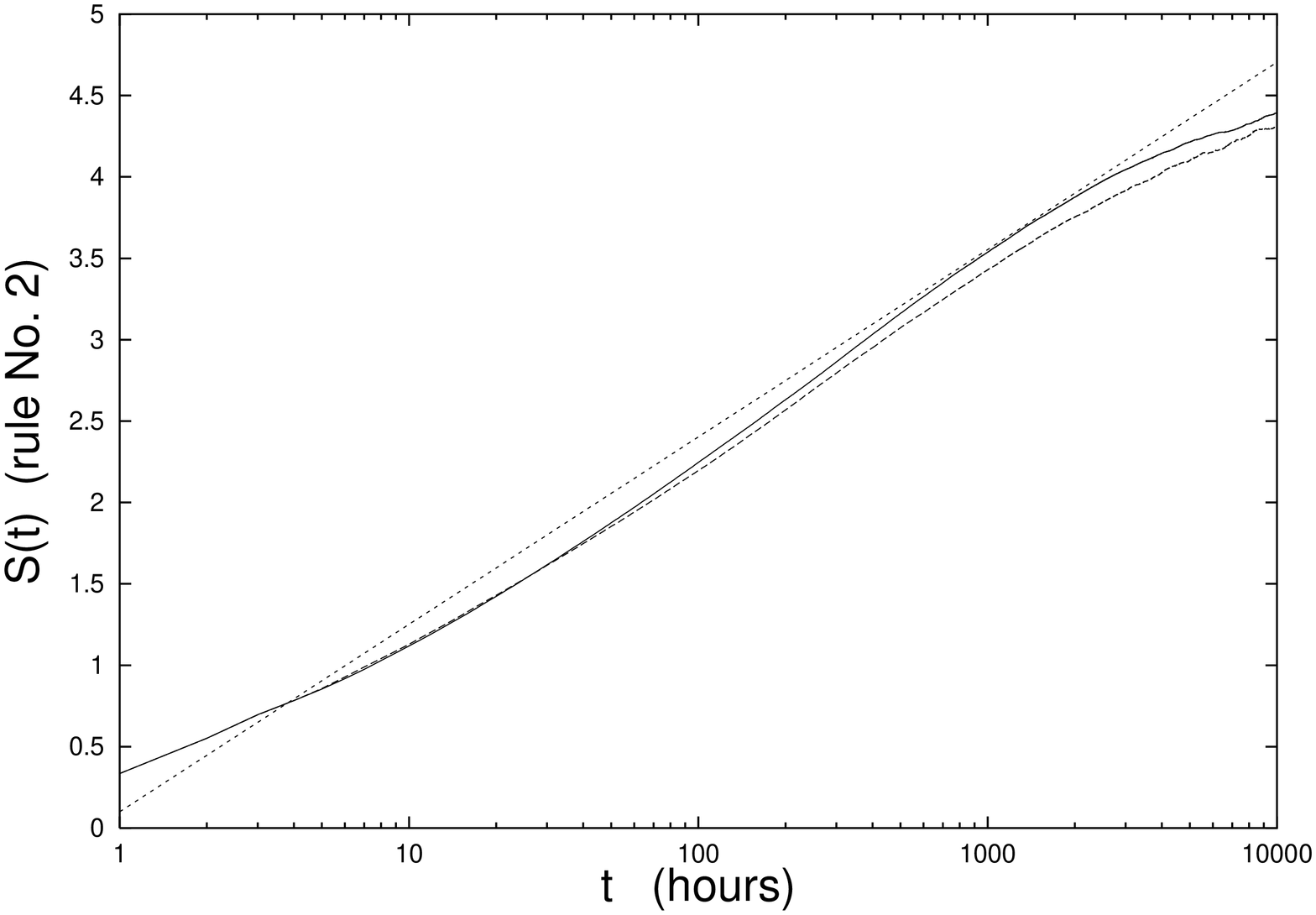, height=16cm,width=12cm,angle=-90}
~\\

\caption{
DE as a function of time according to rule No. 2. The dotted straight line illustrates the slope of entropy increase corresponding to $\mu = 2.144$, $\delta=0.5$, which is the best value of $\mu$ afforded by the analysis of Section V. The dashed line is the DE curve generated by the non-shuffled real data. The solid line is the DE curve generated by the shuffled real data. }
\end{figure}

\newpage

\begin{figure}[h]
\epsfig{file=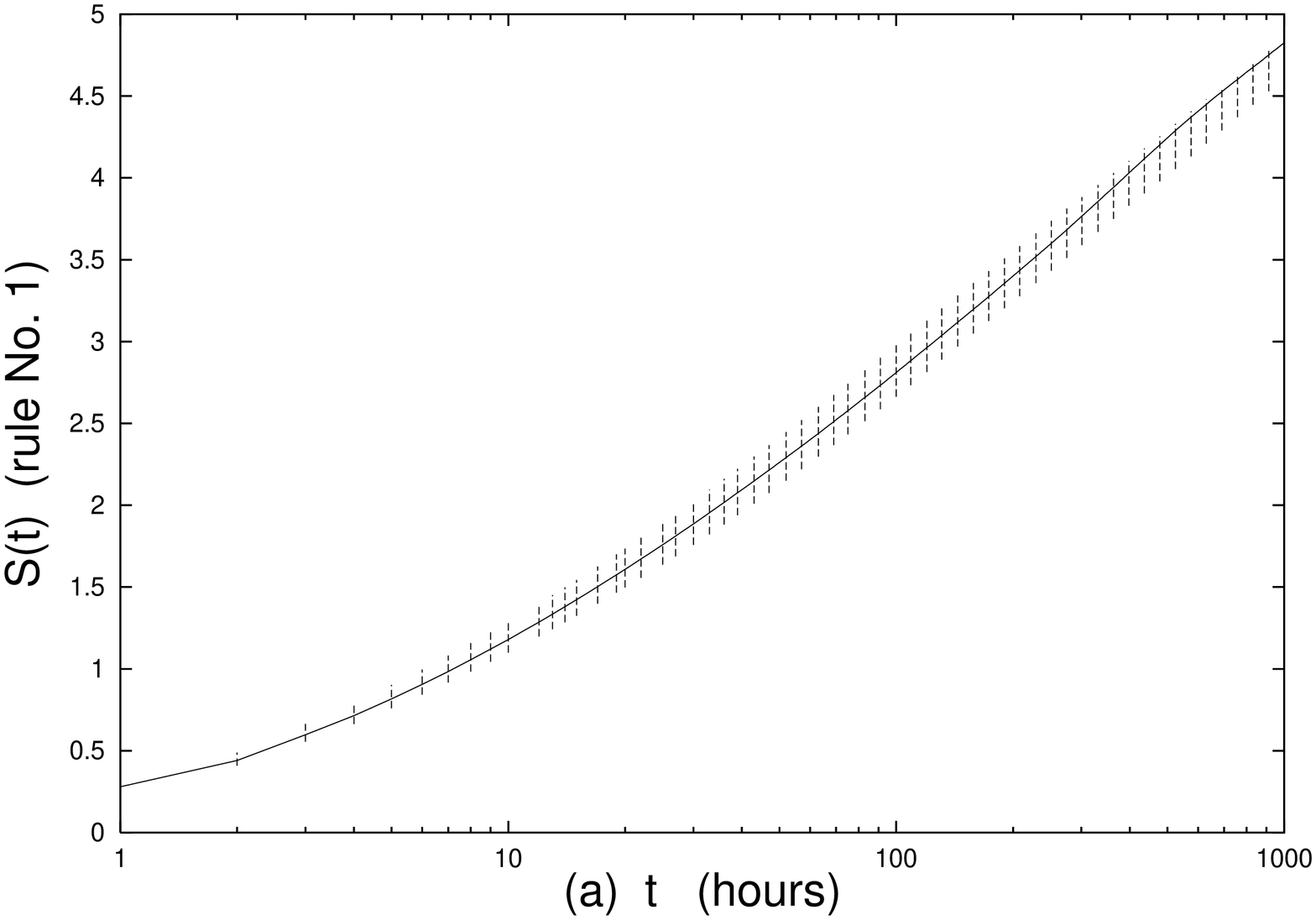, height=16cm,width=8cm,angle=-90}

\epsfig{file=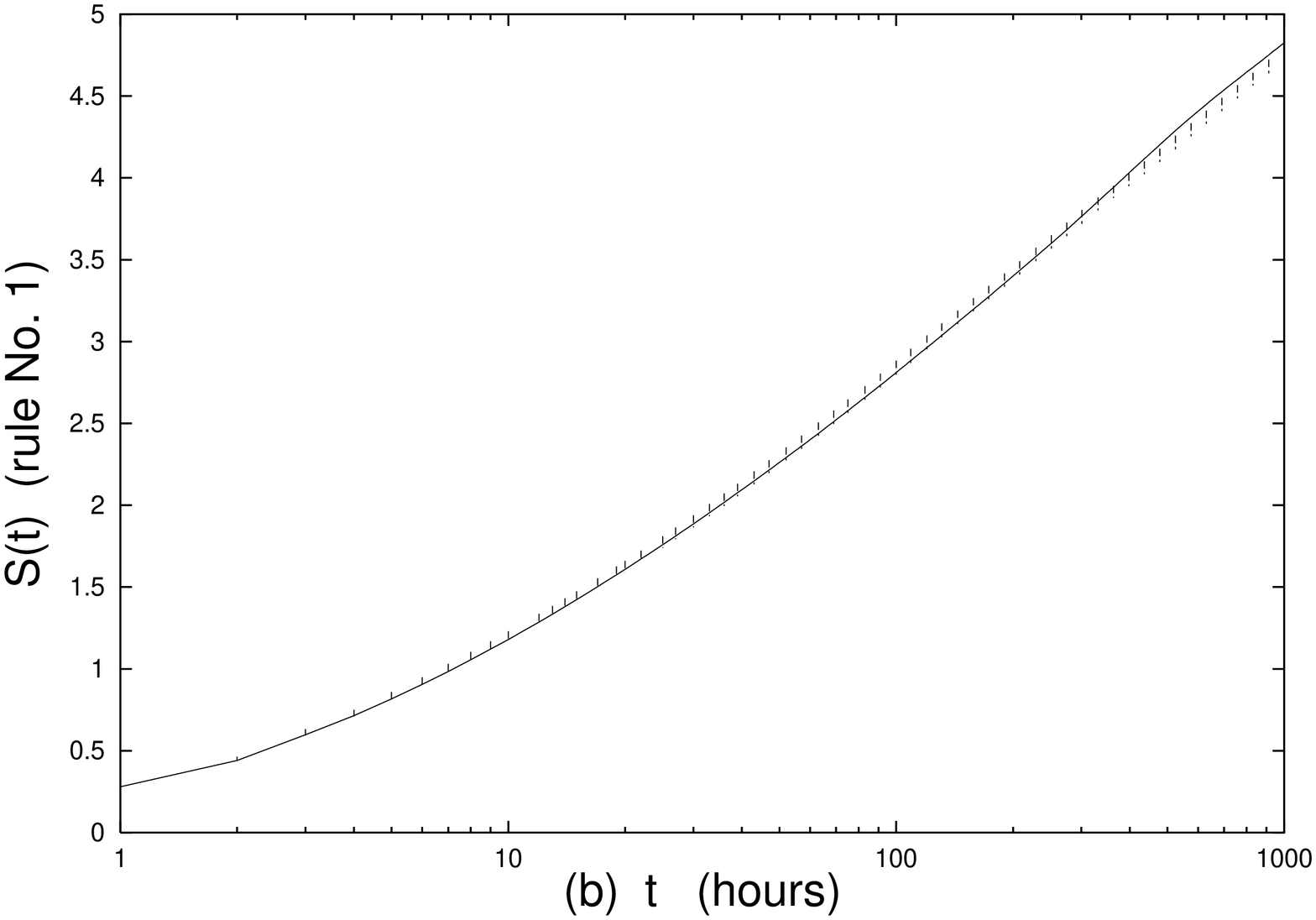, height=16cm,width=8cm,angle=-90}
~\\

\caption{ 
DE as a function of time according to the rule No. 1. The two solid curves denote the DE curve corresponding to the shuffled real data. (a) The vertical bars indicate the changes of the DE curves resulting from the artificial sequences described in the text with $T = 8422$ and $\mu$ moving in the interval [2.094, 2.194].
(b) The vertical bars indicate the changes of the DE curves resulting from artificial sequences described in the text with $\mu = 2.144$, and T moving in the interval [7922, 8922].}
\end{figure}

\newpage

\begin{figure}[h]
\epsfig{file=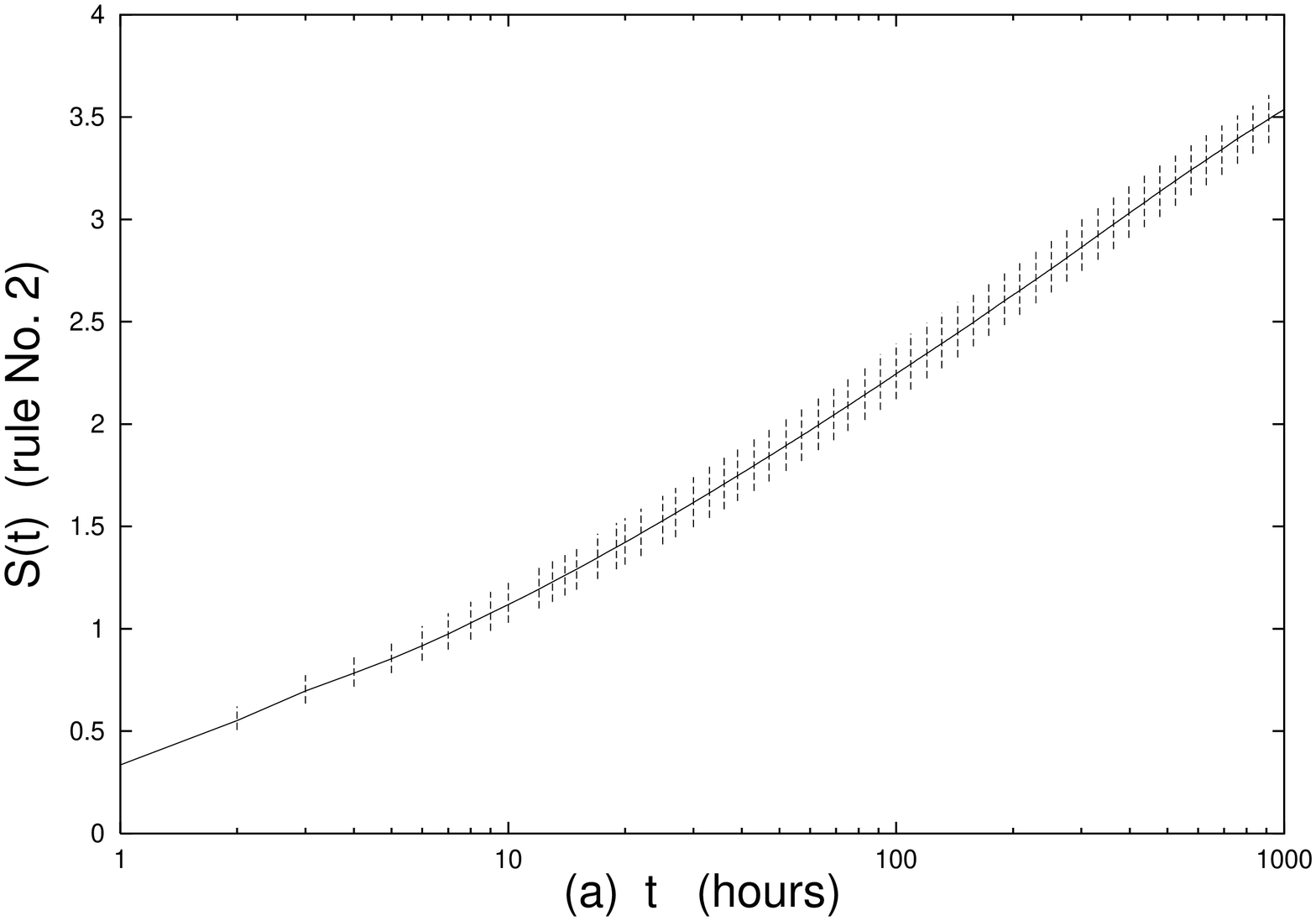, height=16cm,width=8cm,angle=-90}

\epsfig{file=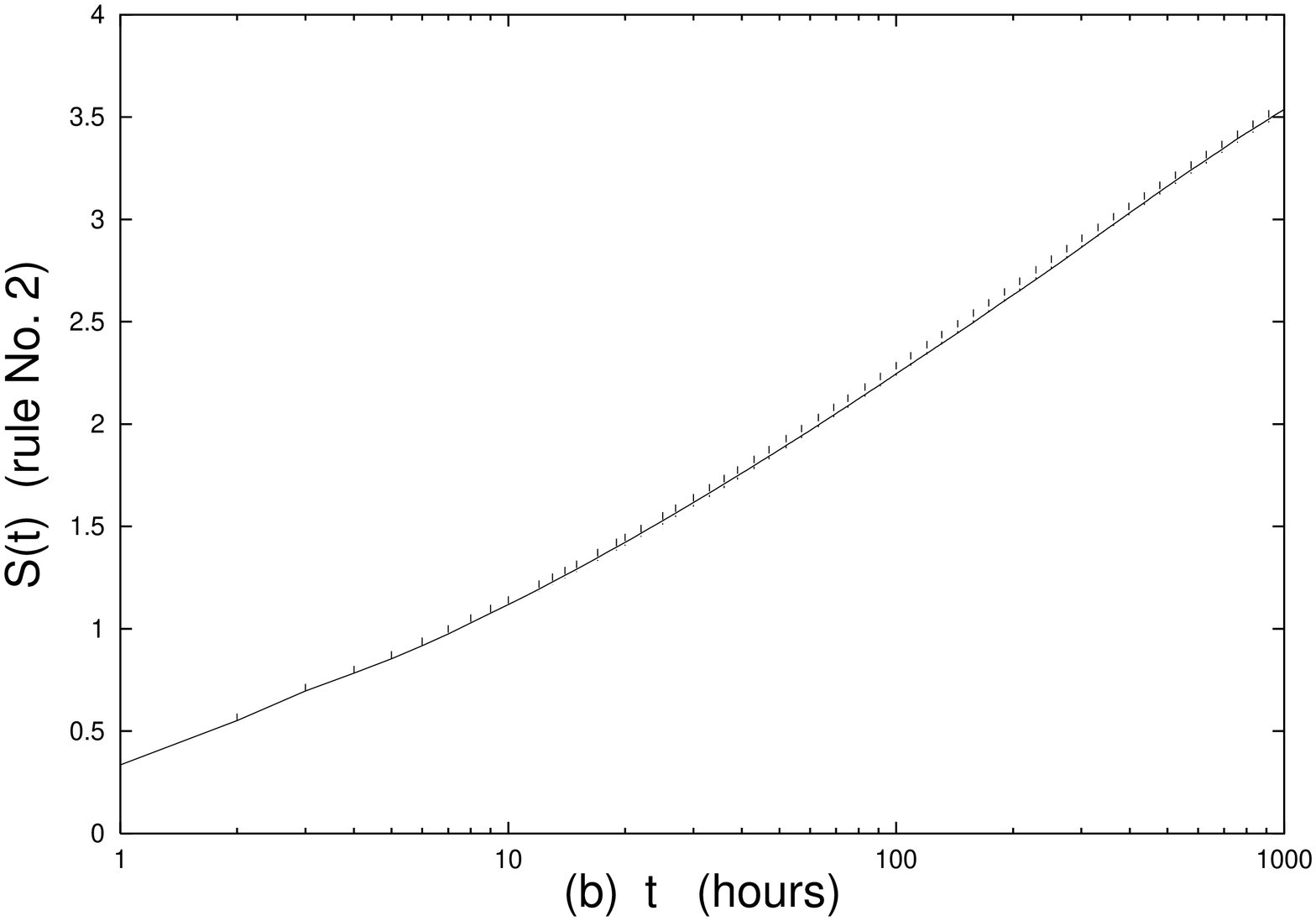, height=16cm,width=8cm,angle=-90}
~\\

\caption{ 
DE as a function of time according to the rule No. 2. The two solid curves denote the DE curve corresponding to the shuffled real data. (a) The vertical bars indicate the changes of the DE curves resulting from the artificial sequences described in the text with $T = 8422$ and $\mu$ moving in the interval [2.094, 2.294]. 
(b) The vertical bars indicate the changes of the DE curves resulting from artificial sequences described in the text with $\mu = 2.144$, and T moving in the interval [7922, 9922]. }
\end{figure}

\newpage

\begin{figure}[h]
\epsfig{file=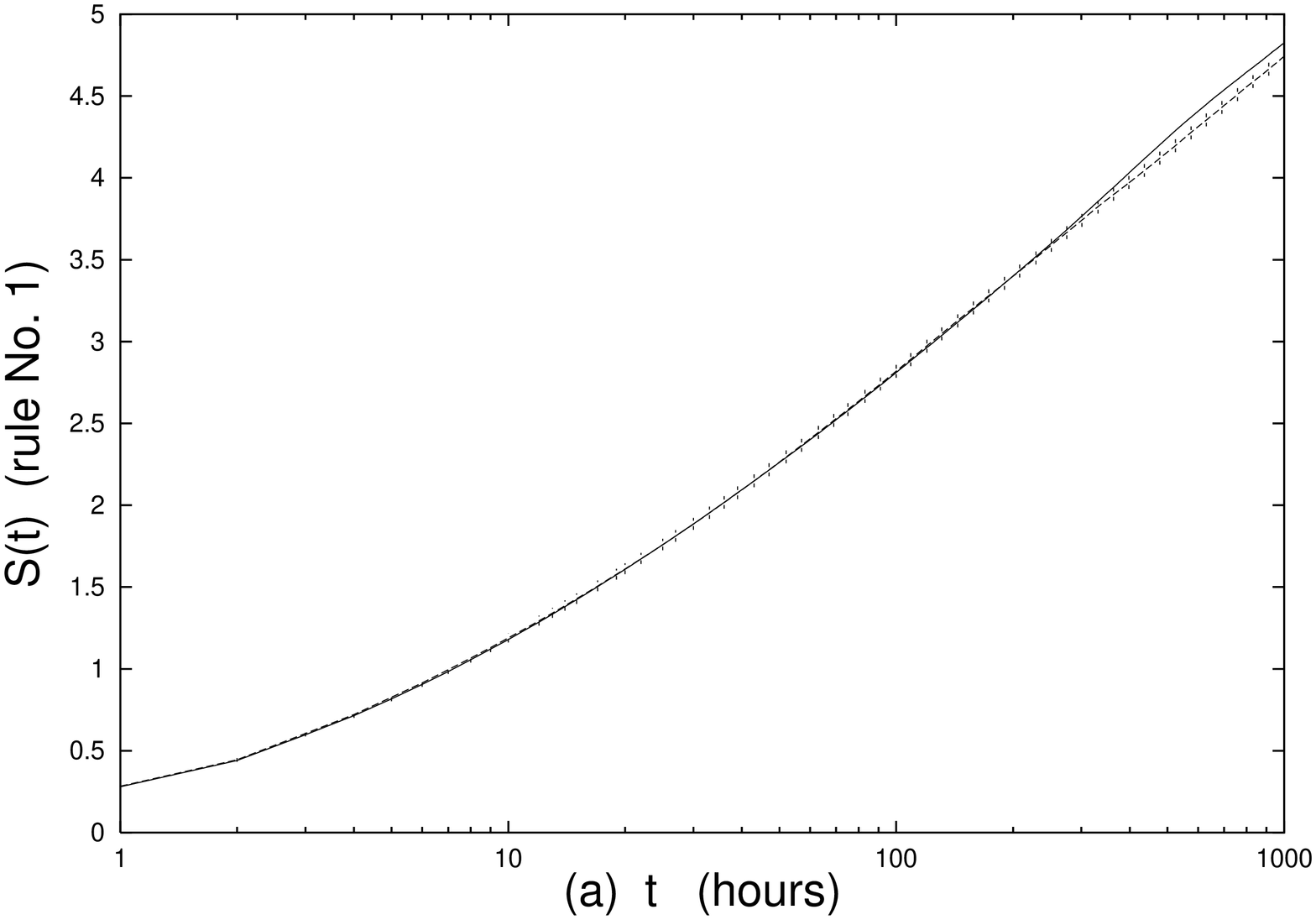, height=16cm,width=8cm,angle=-90}

\epsfig{file=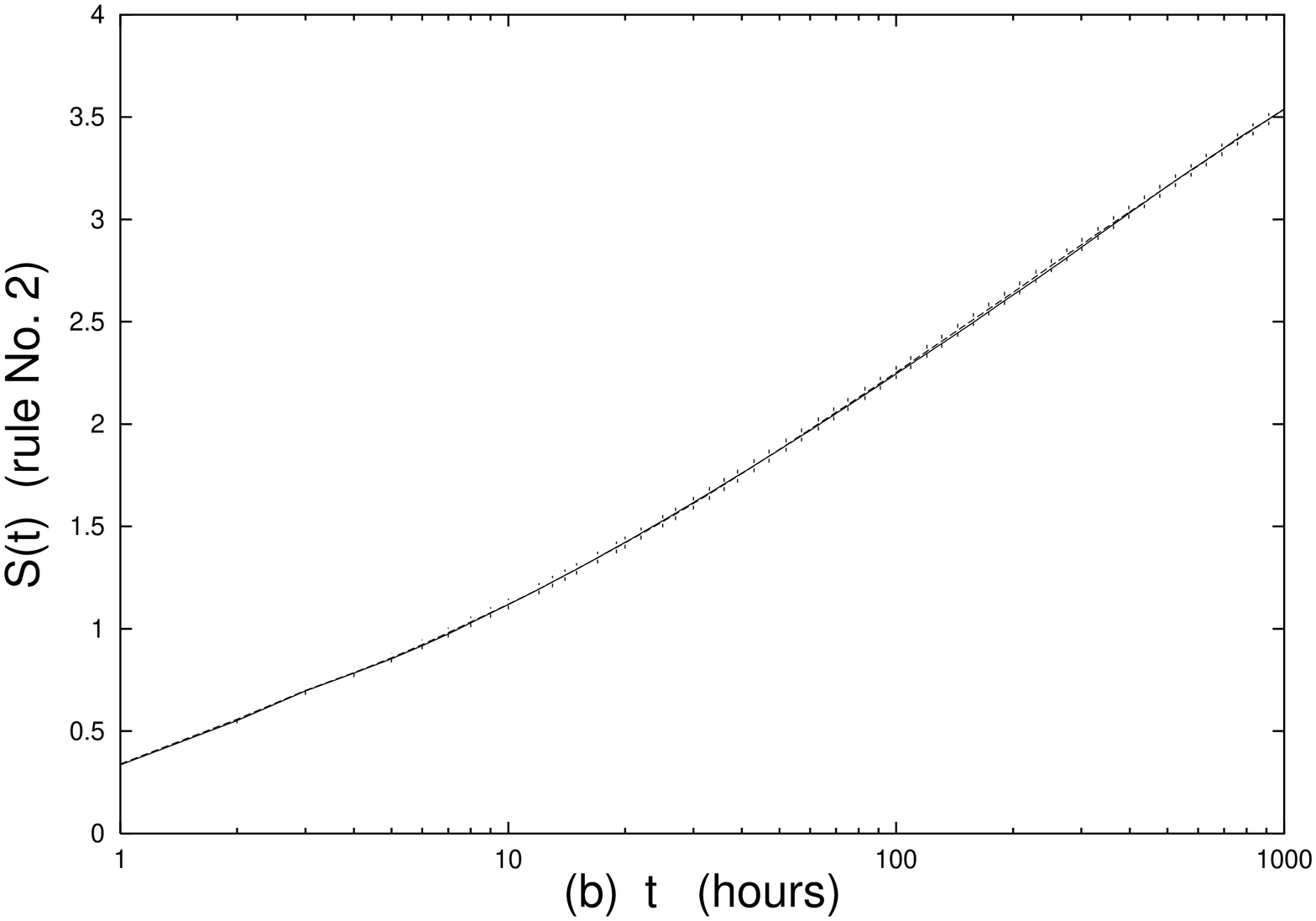, height=16cm,width=8cm,angle=-90}
~\\

\caption{
DE as a function of time. The solid lines denote the DE curve generated by the shuffled real data, and the dashed lines, which almost coincide with the solid lines,  denote the DE curves resulting from the artificial sequence with $\mu = 2.138$ and $T = 8422$.  (a) Rule No. 1. (b) Rule No. 2.  }
\end{figure}

\newpage

\begin{figure}[h]
\epsfig{file=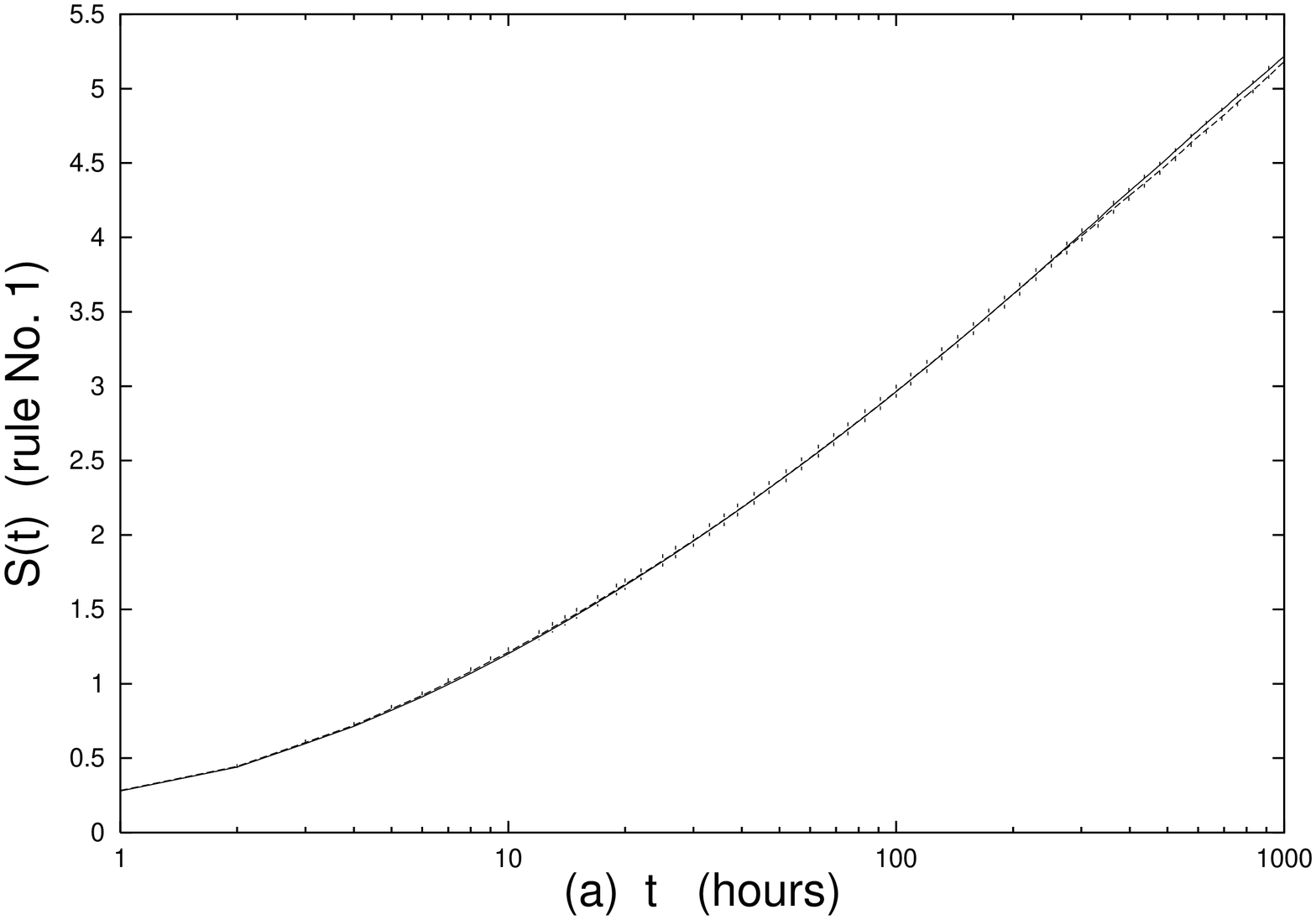, height=16cm,width=8cm,angle=-90}

\epsfig{file=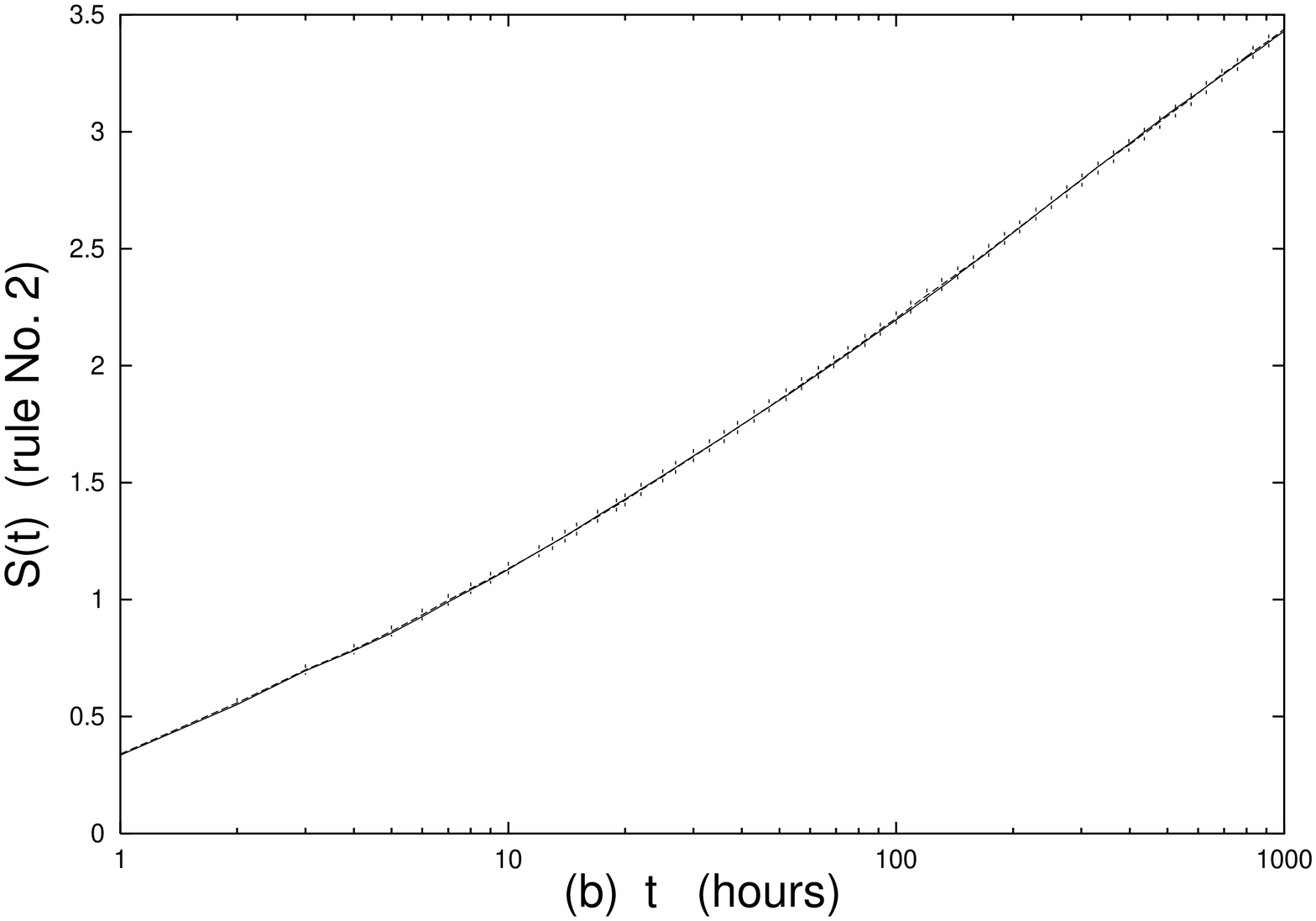, height=16cm,width=8cm,angle=-90}
~\\

\caption{ 
DE as a function of time. The solid lines denote the DE curve generated by the unshuffled real data, and the dashed lines, which almost coincide with the solid lines, denote the DE curves resulting from the artificial sequence with $\mu = 2.138$ and $T = 8422$ with a modulation mimicking the influence of the 11-years solar cycle.  (a) Rule No. 1. (b) Rule No. 2.  }
\end{figure}

\begin{figure}[h]
\epsfig{file=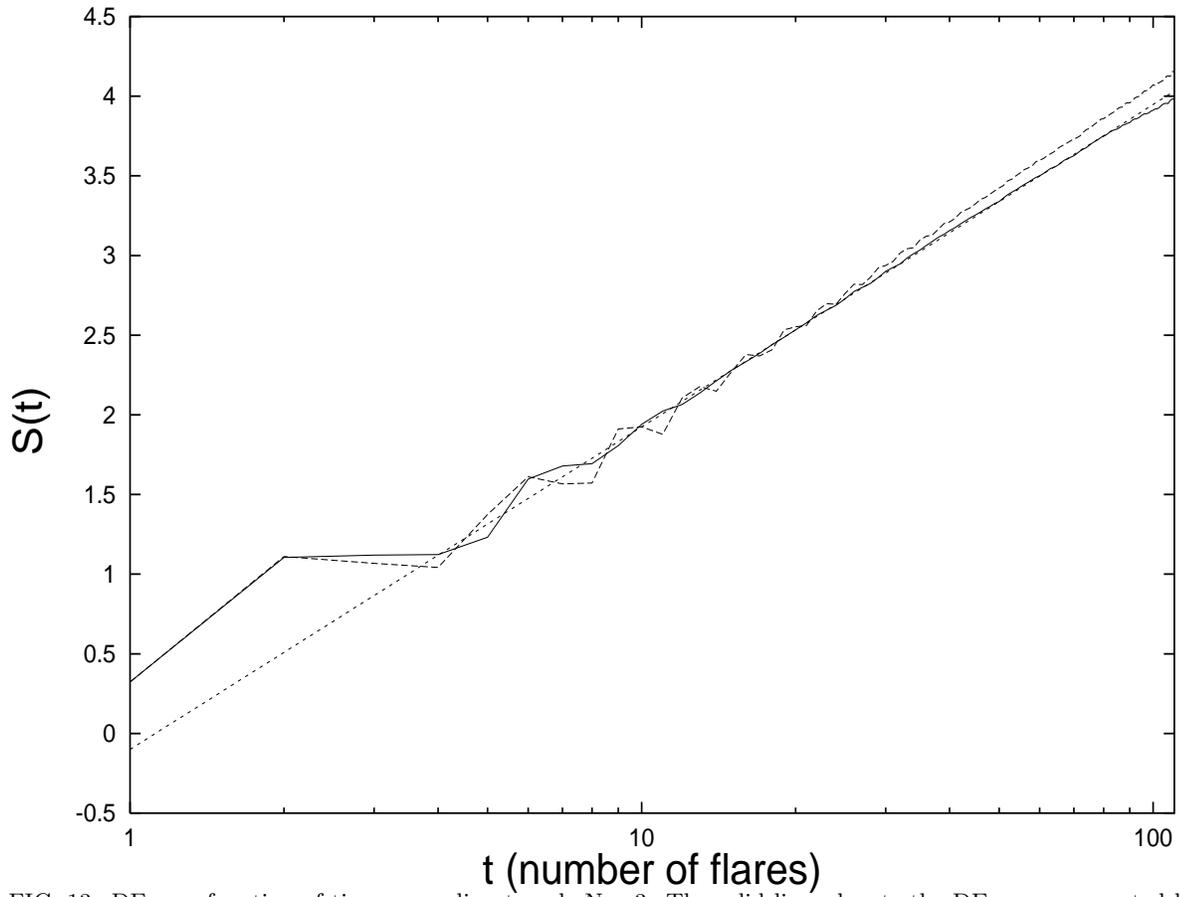, height=16cm,width=12cm,angle=-90}
~\\

\caption{DE as a function of time, according to rule No. 3.  The solid lines denote the DE curve generated by the shuffled real data. The dotted straight line illustrates the slope of entropy increase, $\delta=0.879$, whith corresponds to $\mu = 2.138$. The dashed line  denotes the DE curve resulting from the unshuffled real data. Note the superdiffusion  of the unshuffled real data DE due to the memory in the original signal. 
}
\end{figure}

    \end{document}